\def\year{2021}\relax
\documentclass[letterpaper]{article} %
\usepackage{aaai21}  %
\usepackage{times}  %
\usepackage{helvet} %
\usepackage{courier}  %
\usepackage[hyphens]{url}  %
\usepackage{graphicx} %
\usepackage{xspace}
\usepackage{subcaption}
\usepackage{multirow}
\usepackage{booktabs}
\usepackage{array}
\usepackage{enumitem}
\usepackage{natbib}  %
\usepackage{caption} %
\graphicspath{{graphics/}}

\title{Understanding the Diverging User Trajectories in 
Highly-related Online Communities during the COVID-19 Pandemic}
\author {
    Jason Shuo Zhang,\textsuperscript{\rm 1}
    Brian Keegan, \textsuperscript{\rm 1}
    Qin Lv \textsuperscript{\rm 1}
    Chenhao Tan \textsuperscript{\rm 2} \\
}
\affiliations {
    \textsuperscript{\rm 1} University of Colorado Boulder,
    \textsuperscript{\rm 2} University of Chicago \\
    \{jasonzhang,brian.keegan,qin.lv\}@colorado.edu, chenhao@uchicago.edu
}

\begin{document}

\newcommand{\chenhao}[1]{\textcolor{red}{\textsc{\textbf{[#1 --ct]}}}}
\newcommand{\jason}[1]{\textcolor{blue}{\textsc{\textbf{[#1]}}}}
\newcommand{\para}[1]{\noindent{\bf #1}}
\newcommand{\figref}[1]{Figure~\ref{#1}}
\newcommand{\equationref}[1]{Equation~\ref{#1}}
\newcommand{\secref}[1]{Section~\ref{#1}}
\newcommand{\tableref}[1]{Table~\ref{#1}}

\newcommand{\rchinaflu}{{/r/China\_flu}\xspace}
\newcommand{\rcoronavirus}{{/r/Coronavirus}\xspace}
\newcommand{\rchinaflus}{{/r/China\_flu's}\xspace}
\newcommand{\rcoronaviruss}{{/r/Coronavirus'}\xspace}
\newcommand{\rconspiracy}{{/r/conspiracy}\xspace}
\newcommand{\rconservative}{{/r/Conservative}\xspace}
\newcommand{\rpolitics}{{/r/politics}\xspace}
\newcommand{\communityname}[1]{{#1}\xspace}
\newcommand{\theendday}{September 30th\xspace}
\newcommand{\fightin}{Fightin' Words model\xspace}

\maketitle

\begin{abstract}
As the COVID-19 pandemic is disrupting life worldwide, related online communities are popping up.
In particular, two ``new'' communities, \rchinaflu and \rcoronavirus, emerged on Reddit and have been dedicated to COVID-related discussions from the very beginning of this pandemic.
With \rcoronavirus promoted as the official community on Reddit, it remains an open question how users choose between these two highly-related communities.

In this paper, 
we characterize user trajectories in these two communities from the beginning of COVID-19 to the end of September 2020.
We show that new users of \rchinaflu and \rcoronavirus were similar from January to March. 
After that, their differences steadily increase, evidenced by both language distance and membership prediction, as the pandemic continues to unfold.
Furthermore, users who started at \rchinaflu from January to March were more likely to leave,
while those who started in later months tend to remain highly ``loyal''.
To understand this difference, we develop a movement analysis framework to understand membership changes in these two communities and identify a significant proportion of 
\rchinaflu members (around 50\%) that moved to \rcoronavirus in February. 
This movement turns out to be highly predictable based on other subreddits that users were previously active in.
Our work demonstrates how two highly-related communities emerge and develop their own identity in a crisis, and highlights the important role of existing communities in understanding such an emergence.
\end{abstract}

\section{Introduction}
\begin{figure*}[t]
    \center
    \begin{subfigure}[t]{0.4\textwidth}
        \includegraphics[width=0.9\textwidth]{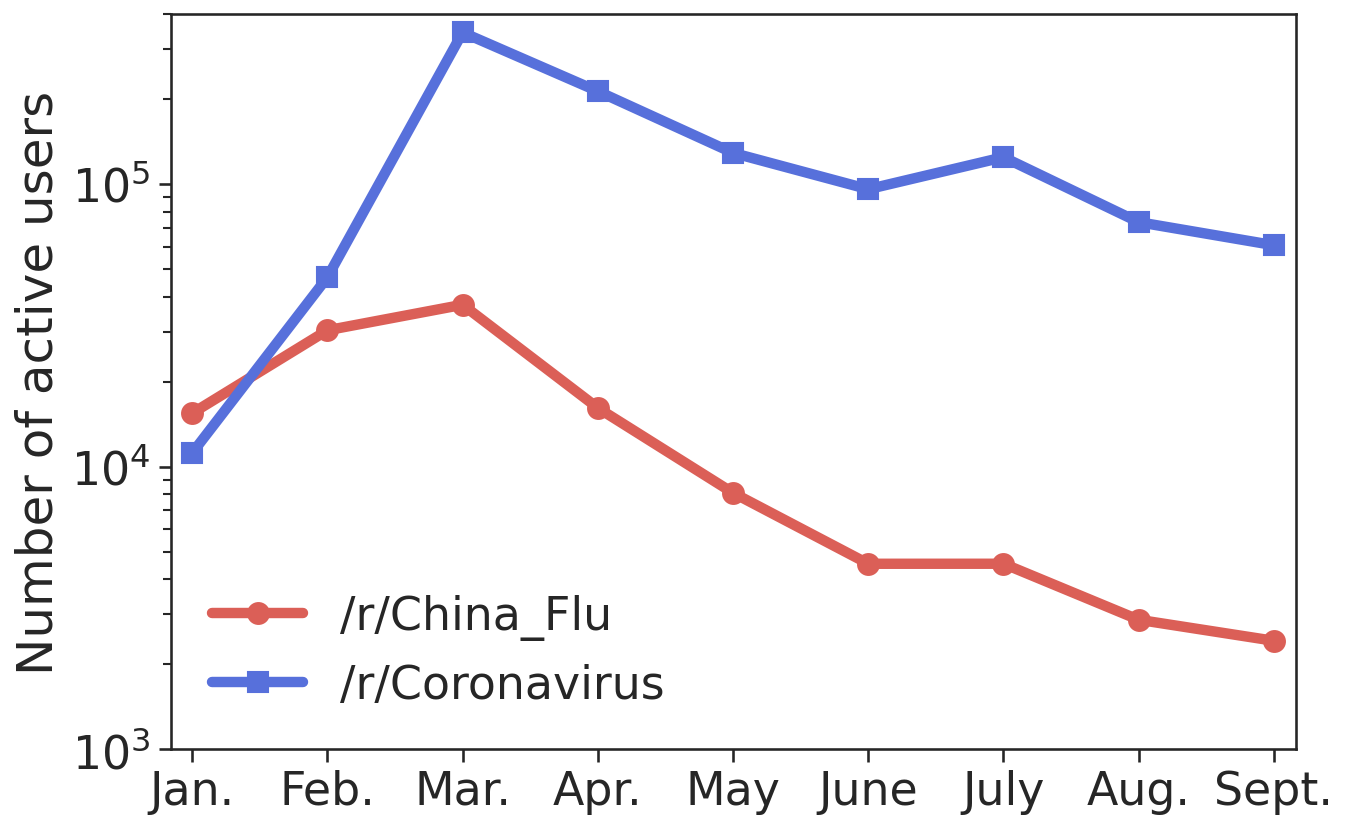}
        \caption{Number of active users per month.}
        \label{fig:num_users}
    \end{subfigure}
    \begin{subfigure}[t]{0.4\textwidth}
        \includegraphics[width=0.9\textwidth]{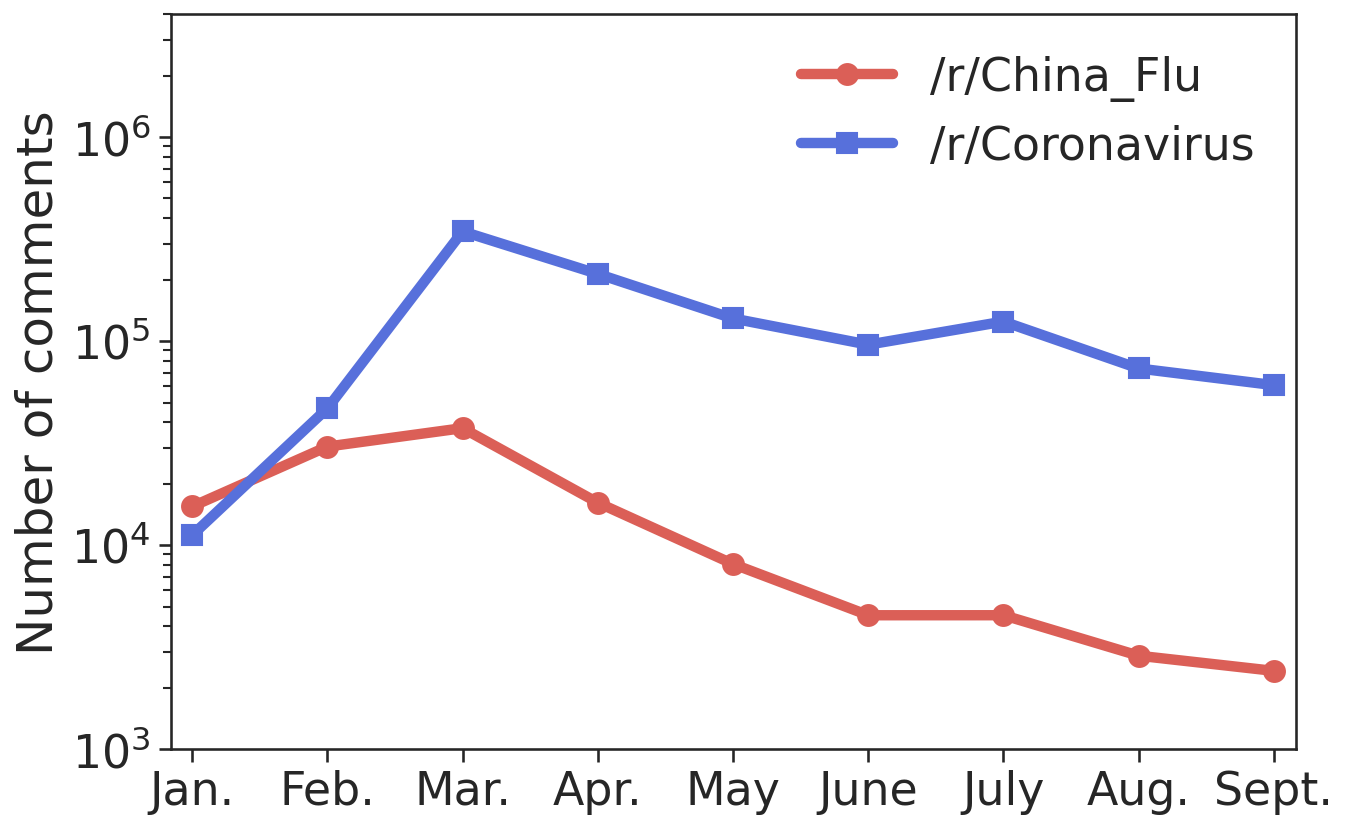}
        \caption{Number of comments per month.}
        \label{fig:num_comments}
    \end{subfigure}
    \caption{
     The number of active users and comments per month, respectively, in \rchinaflu and \rcoronavirus on a {\em log} scale. 
    In January and February, user activity was at a similar level between these two 
    subreddits. After that, \rcoronavirus became much more popular than \rchinaflu.
    }
    \label{fig:total_num}
\end{figure*}
In December 2019, a novel coronavirus strain (SARS-CoV-2) emerged in Wuhan, China.
The disease (COVID-19) quickly spread all over the world and led the World Health Organization (WHO) to
declare a pandemic \cite{unfolded}. 
By the end of September 2020, 189 countries/regions reported over 34 million positive cases 
and 1.0 million deaths \cite{dong2020interactive}.
The pandemic is not only exhausting public health resources but also causing social and economic 
disruption at an unprecedented speed and scale. 
Social media are critical for people to access information and share experiences during this pandemic.
Take Reddit as an example.
User-created subreddits (communities) 
have popped up for discussions on coronavirus and attracted millions of subscribers.
Health professionals~\cite{rcorona}, essential workers~\cite{grocery}, and recovered patients~\cite{positive} are reportedly using Reddit as a primary source to share 
information.
Despite the importance of these communities,
it remains an open question as to how these highly-related communities emerge and form their identity in the context of a crisis.

In this work, we 
focus on two 
highly-related
communities on Reddit, \rchinaflu and \rcoronavirus.
Both communities are for {\em general} discussions about COVID-19 and were ``founded'' 
at the very beginning of the pandemic. 
There were more active users in \rchinaflu than in \rcoronavirus 
in January 2020 (\figref{fig:total_num}). 
After that, \rcoronavirus exploded, as the platform made it the official community 
for COVID-19 on February 17th (\figref{fig:feb17}).
In comparison, user activity in \rchinaflu has gradually decreased.
The emergence of these two communities during a once-in-a-lifetime pandemic 
provides a unique opportunity for understanding the formation of communities during a crisis, bridging the literature on highly-related communities~\citep{hessel2016science,tan2018tracing,zhu2014selecting,zhang2018we,zhu2014impact,waller2019generalists} and crisis informatics~\citep{palen2007crisis,palen2009crisis,palen2016crisis,hagen2018crisis,reuter2018fifteen,reuter2018social,maas2019facebook}.

We take a user-centered perspective, as communities are ultimately determined by their members.
A key observation is that users in these two communities
were similar from January to March and then diverge.
To further understand this divergence,
we examine two main questions:
1) Which users chose to start at \rchinaflu vs. \rcoronavirus:
We find that new users who started at \rchinaflu were already different 
from those who started at \rcoronavirus in their use of language before joining either community, and
this difference has grown over time.
2) How users move between these two communities:
We find that a significant proportion of \rchinaflu (around 50\%) members moved to 
\rcoronavirus in February.
Moreover, this movement is highly predictable based on other subreddits that users were previously active in.

\para{Organization and highlights.}
We provide a detailed overview of \rchinaflu and \rcoronavirus 
after discussing related work.
We demonstrate how user activity is connected with the important dates of this pandemic, 
including February 17th, when Reddit made \rcoronavirus the official subreddit for COVID-19, 
and March 11th, when the WHO declared COVID-19 a pandemic.
We show that both communities share similar founders who tend to be newsreaders or survivalists.
In March, more than half of the \rchinaflu users were also active in \rcoronavirus. 
After that month, the overlap decreases over time.

Inspired by the separation of active users in these two communities,
we compare the differences between cohorts of {\em new} users who chose to start at \rchinaflu vs. at \rcoronavirus, 
grouped by their starting month.
Our results show that the language distance between users who started at \rchinaflu or \rcoronavirus 
was small from January to March. 
After that, the language distance goes up, indicating an increasing distinction between users who chose to start at \rchinaflu vs. \rcoronavirus.
We observe a similar trend in a prediction task:
Users' previous activity on Reddit can be used to predict which community they chose to start at, and 
the accuracy is better in later months than in earlier ones.
These observations indicate that the separation between \rchinaflu and \rcoronavirus users 
has widened as the pandemic unfolds.

In addition to the choice of where to start, 
we further examine whether users move between these communities.
We find that users who started at \rchinaflu from January to March reduced their activity in this community over time, 
while those in later months tend to be highly ``loyal''.
To further understand the movement between these two communities, 
we develop a framework to identify membership changes.
Our framework shows that around 50\% of \rchinaflu members moved to \rcoronavirus
in February, when \rcoronavirus became the official community for COVID-19 on Reddit.
This ratio is robust, even if we apply different membership definitions.
In comparison, users rarely moved in the reverse direction.
Moreover, we demonstrate that these users' movements from \rchinaflu to \rcoronavirus 
are highly predictable based on other subreddits that users were previously active in.
Donald Trump supporters, conspiracy theorists, and survivalists were more likely to stay
in \rchinaflu. In contrast, Bernie Sanders (a U.S. Democratic presidential candidate) supporters 
and science enthusiasts were more likely to leave \rchinaflu.

We offer concluding discussions in the end.
Our work demonstrates the emerging process of two highly-related communities in a crisis 
through the perspective of their members.
We show that these two communities resemble each other in the beginning and then gradually diverge.
Despite the dominance of \rcoronavirus, \rchinaflu forms its own identity and can still attract users 
with high loyalty as the pandemic unfolds.
We also highlight the critical role of existing communities in this process.

\section{Related Work}
\label{sec:related}

We review the literature in two areas that are most relevant to our work:
highly-related communities and crisis informatics.

\subsection{Highly-related Communities \& Community Genealogy}
When social media platforms give users the freedom to form interest groups, 
a series of highly-related communities can pop up. 
For example, during the 2016 election, a battery of Trump-related communities %
are created on Reddit, 
such as \communityname{/r/The\_Donald}, \communityname{/r/AskThe\_Donald}, and 
\communityname{/r/AskTrumpSupporters}. 
The creation, development, and lifecycle of highly-related communities have drawn considerable interest
in the research community.
The first line of work compares the characteristics of highly-related communities.
\citet{hessel2016science} investigate the interactions between highly-related Reddit communities
and identify patterns of affixes being used in their names. 
The work by \citet{zhang2018we,zhang2019intergroup} focuses on online NBA fan communities and
analyzes how fans of different teams talk with each other and react to team performance.
The second line of research studies the impact of membership overlap between highly-related communities.
Haiyi Zhu, Robert Kraut, and collaborators investigate shared membership in online communities
on a variety of platforms and propose strategies for long-term community survival 
\cite{zhu2014impact,zhu2013effectiveness,kraut2012building,zhu2014selecting}.
The final line provides a global overview of how users move through the space of communities and how
new communities emerge from the old ones.
For example, \citet{tan2015all} use several temporal features across communities 
to predict users' activity levels and departure from Reddit.
To understand how new communities are developed from old ones,
\citet{tan2018tracing} proposes a computational approach for building genealogy graphs
between communities.

\subsection{Crisis Informatics \& Sociology of Disaster}
Another closely related line of work is crisis informatics \cite{palen2016crisis}. 
Understanding disaster events and their 
impacts 
is a critical topic of societal relevance~\cite{undisaster}. 
In the early days of crisis informatics research, qualitative methods, such as descriptive surveys and interviews, 
were the main sources of data collection \cite{palen2007crisis,suttonbackchannels,palen2009crisis}. 
Recently, following the advancement of the Internet and mobile technology, social media have played 
a critical role in the flow of public information.
A growing percentage of citizens frequently turn to these platforms for emergency updates \cite{lachlan2016social,rene2016influence}.
To effectively analyze information on mass media,  
computational approaches have been widely adopted in crisis informatics. 
For example, \citet{vieweg2010microblogging} analyze public responses to two disaster events, the Red River
Floods and the Oklahoma Grassfires, using Twitter communications.
\citet{hagen2018crisis} propose a network analysis approach to identify a number of distinct communities
and influential actors using Zika-related tweets. 
Using Hurricane Sandy as a study subject,
\citet{stewart2016dynamic} characterize how citizens utilize social media to redistribute emergency updates 
and connect with family and friends.

Despite tremendous effort in these two areas, the phenomenon of highly-related communities emerging 
during a crisis is understudied.
The substantial impact caused by COVID-19 provides an opportunity to understand this phenomenon.
To the best of our knowledge, our work represents the first attempt towards this direction by unpacking the emerging process of two highly-related communities.

\para{Recent studies on COVID-19.} Meanwhile, there are recent studies that examine the public response to the COVID-19 pandemic \cite{van2020using,yin2021coevolution}.
In particular, sentiment and language usage 
in
COVID-related conversations on social media are investigated from various angles 
\cite{budhwani2020creating,chen2020eyes,pei2020coronavirus,lyu2020sense}. 
In this work, we also compare the language usage between users in \rchinaflu and \rcoronavirus to investigate
the separation of these two communities as the pandemic unfolds.

\section{An Overview of \\ {\rchinaflu} and {\rcoronavirus}}
\label{sec:overview}

Our main dataset is drawn from Reddit,\footnote{\url{https://www.reddit.com/}.} 
a community-driven forum for discussion, news consumption, 
and content rating. 
It was founded in 2005 and became the 21st most popular website globally in May 2020 \cite{reddittop}. 
There are tens of thousands of communities (known as subreddits) on Reddit 
dedicated to a wide
variety of topics. 
Users can submit, comment on, upvote, and downvote content in each subreddit. 

Since the outbreak of COVID-19 in January 2020, a series of COVID-related communities
have emerged and drawn substantial public attention. 
In this study, we focus on two communities, \rchinaflu\footnote{
On the front page of \rchinaflu (\url{https://www.reddit.com/r/China_Flu/}), it explains, ``The name \rchinaflu was created at a time when SARS-CoV-2 had not been named and was only affecting China. Subreddit names cannot be changed after they are created.''} and \rcoronavirus,\footnote{\url{https://www.reddit.com/t/coronavirus/}.} which have been dedicated 
to general COVID-related discussions from the very beginning of this pandemic. 
Other COVID-related communities tend to focus on subtopics such as scientific discussions (e.g., \communityname{/r/COVID19}) and specific regions (e.g., \communityname{/r/CanadaCoronavirus}).

\rchinaflu was founded on January 20th, 2020, when news about the first breakout 
of COVID-19 emerged in Wuhan, China.  
\rcoronavirus was founded on May 3rd, 2013, but remained inactive most of the time. 
The first post in \rcoronavirus since 2017 was about COVID-19 on January 20th, 2020. 
Using the pushshift.io website \cite{baumgartner2020pushshift}, we collect all the comments 
submitted to these two subreddits from January 20th to \theendday in 2020.
\tableref{tab:stats} gives summary statistics about these two subreddits.
In the rest of the paper, we focus on analyzing users' comments since posts are usually 
direct links to news articles and more formal, 
thus
not comparable to comments. 

\begin{figure}[t]
    \centering
    \includegraphics[width=0.45\textwidth]{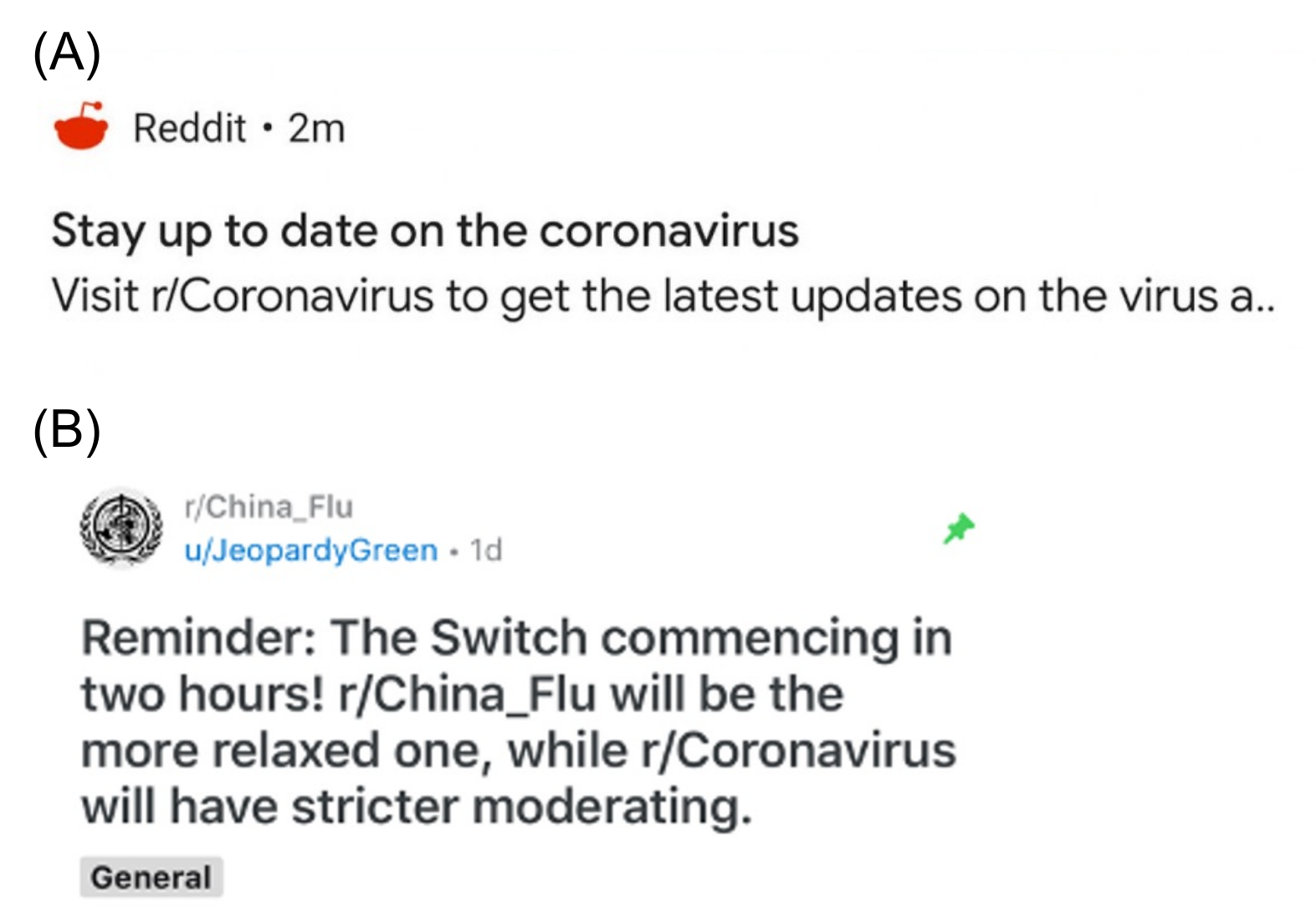}
    \caption{On February 17th, Reddit decided to make \rcoronavirus the official community 
    for COVID-19. All the comments submitted to \rcoronavirus are heavily moderated.
    Meanwhile, the platform allows more relaxed discussions in \rchinaflu.
    }
    \label{fig:feb17}
\end{figure}

\begin{table}[t]
\centering
\begin{tabular}{lrrr}
\toprule
         & \#posts & \#comments & \#users \\
\midrule
\rchinaflu    & 81K  & 1.3M  &   76K  \\ 
\rcoronavirus    & 312K    & 7.9M     &    664K \\ 
\bottomrule
\end{tabular}
\caption{Dataset statistics as of September 30th, 2020. Here \#users refers to the number of
unique users who have posted/commented in that subreddit.}
\label{tab:stats}
\end{table}

\begin{table}[t]
\centering
\small
\begin{tabular}{lr}
\toprule
                Subreddit & prop. \\
\midrule
        /r/collapse &  28\% \\
       /r/worldnews &  18\% \\
            /r/news &  14\% \\
       /r/AskReddit &  10\% \\
   /r/todayilearned &  10\% \\
        /r/preppers &  10\% \\
      /r/conspiracy &   8\% \\
             /r/aww &   6\% \\
     /r/environment &   6\% \\
 /r/CollapseSupport &   6\% \\
\bottomrule
\end{tabular}
\quad
\begin{tabular}{lr}
\toprule
                    Subreddit & prop. \\
\midrule
                /r/news &   8\% \\
           /r/worldnews &   4\% \\
           /r/AskReddit &   4\% \\
                 /r/MMA &   4\% \\
   \scriptsize{/r/interestingasfuck} &   4\% \\
              /r/AskMen &   4\% \\
 \scriptsize{/r/perfectlycutscreams} &   2\% \\
            /r/collapse &   2\% \\
            /r/preppers &   2\% \\
                /r/nCoV &   2\% \\
\bottomrule
\end{tabular}
    \caption{The top-10 subreddits that founders of \rchinaflu (left) and \rcoronavirus (right) had commented on 30 days before
    they joined these two subreddits.
    Here we define founders of each community as the first 50 users who commented in that community.
    Prop. indicates the percentage of founders that were active in the parent community.
    }
\label{tab:parents}
\end{table}

\subsection{Important Dates}
On February 11th, the WHO named the new coronavirus disease ``COVID-19''. 
It was specifically named this way to avoid calling it the China virus or the Wuhan virus \cite{namecovid}.
On February 17th, the Reddit platform decided to make \rcoronavirus the official community 
for COVID-related updates. 
Users who searched for COVID-related keywords have since been recommended to check out 
\rcoronavirus (\figref{fig:feb17}(A)). 
The subreddit is described as a place for high-quality discussions. 
All the posts and comments submitted are strictly moderated.
Meanwhile, the platform allows for more relaxed discussions in \rchinaflu (\figref{fig:feb17}(B)).
This policy shift is reflected by user activity in these two subreddits.
As illustrated in \figref{fig:total_num}, the monthly number of posts and comments in these
two communities was similar in January and February. 
After that, \rcoronavirus became much more popular than \rchinaflu.

On March 11th, the WHO declared COVID-19 a pandemic \cite{unfolded}.
The number of comments generated in \rchinaflu and \rcoronavirus peaked that month.
User activity in both communities started to drop after that month. 
This may be due to people's fatigue of COVID-related topics.
According to a study released by the Pew Research Center at the end of April,
71\% of Americans said they needed to take a break from news about the coronavirus, 
and 43\% said they felt worse emotionally as a result of following updates~\cite{pewbreaks}.
The user activity in \rcoronavirus bounced back a bit in July, likely due to the second wave 
of coronavirus infections in the U.S. and many other countries~\cite{unfolded}, 
while user activity in \rchinaflu kept going down.

To summarize, the explosion of \rchinaflu and \rcoronavirus happened from January to March, 
when the virus began to spread worldwide. 
Therefore, the period from January to March might be especially interesting 
for understanding the emergence of these two communities.

\subsection{Membership in \rchinaflu and \rcoronavirus}

We present exploratory analyses of membership in these two communities.
We start by looking at the early users (``founders'') in each community and then examine the overlap of active users in these two communities over time.

\para{Founders of \rchinaflu and \rcoronavirus.} A straightforward way to explore the beginning stage of these two new communities is 
to understand where the founders of \rchinaflu and \rcoronavirus came from.
\tableref{tab:parents} shows the top-10 parents of these two communities.
These top-10 parents are ranked based on each community's first 50 commenters'
activity on Reddit 30 days before joining \rchinaflu or \rcoronavirus.
Interestingly, the founders of \rchinaflu and \rcoronavirus were both
active in \communityname{/r/collapse} and \communityname{/r/preppers}, 
two subreddits that attract survivalists
to discuss the potential collapse of global civilization. 
It suggests that people who paid attention to this virus at the very beginning
tend to worry about the break down of our society in general \cite{collapse}. 
The dominance of \communityname{/r/collapse} is more salient in \rchinaflu than \rcoronavirus.
Unsurprisingly, early members of \rchinaflu and \rcoronavirus were also active in
news-related subreddits (\communityname{/r/worldnews} and \communityname{/r/news}).
This explains why they were the earliest to be aware of this new disease.
Moreover, \communityname{/r/conspiracy} is one of the top-10 subreddits for \rchinaflu,
indicating that COVID-19 topics attract conspiracy theorists \cite{conspiracy,conspiracyny}.

\begin{figure}
    \center
    \includegraphics[width=0.4\textwidth]{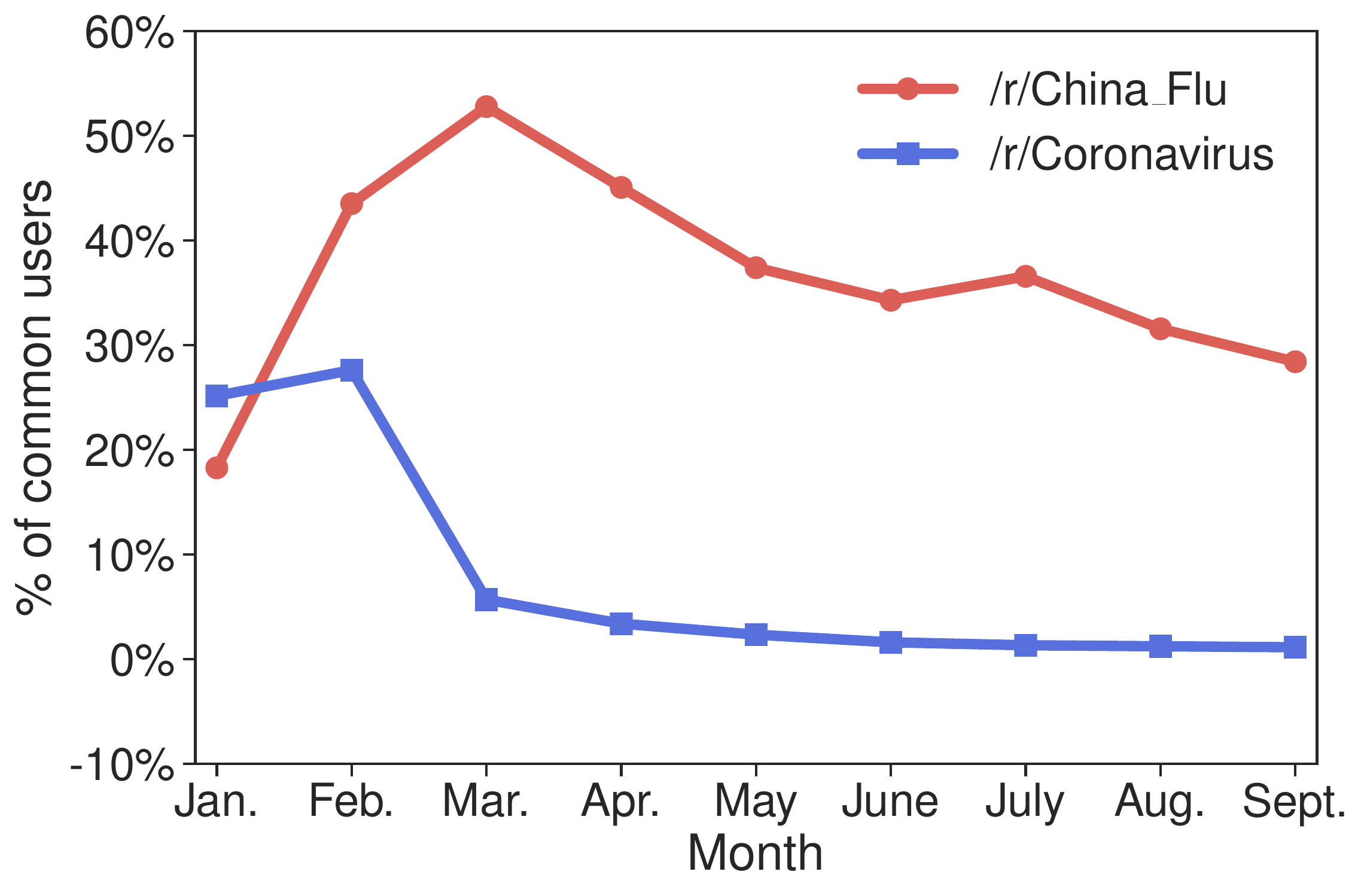}
        \caption{The proportion of monthly overlapping users between \rchinaflu and \rcoronavirus.
    The overlap is measured by the proportion of commenters that are active in both communities.
    }
    \label{fig:common}
\end{figure}

\para{The monthly overlap between \rchinaflu and \rcoronavirus.}
Next, we examine the overlap between these two communities as the pandemic unfolds.
We define the overlap as the proportion of commenters in one community that also commented in the other
community in the same month (see \figref{fig:common}). 
In March, more than half of the \rchinaflu users were also active in \rcoronavirus.
After that month, the overlap goes down over time. 
This trend indicates that the separation between \rchinaflu and \rcoronavirus 
has widened as the pandemic unfolds.
As \figref{fig:total_num} shows, \rcoronavirus has a much bigger user base since March.
The fraction of its users who are also active in \rchinaflu is low.

\subsection{Mention of ``china''}
\label{sec:china_mention}

\begin{figure}
    \center
    \includegraphics[width=0.35\textwidth]{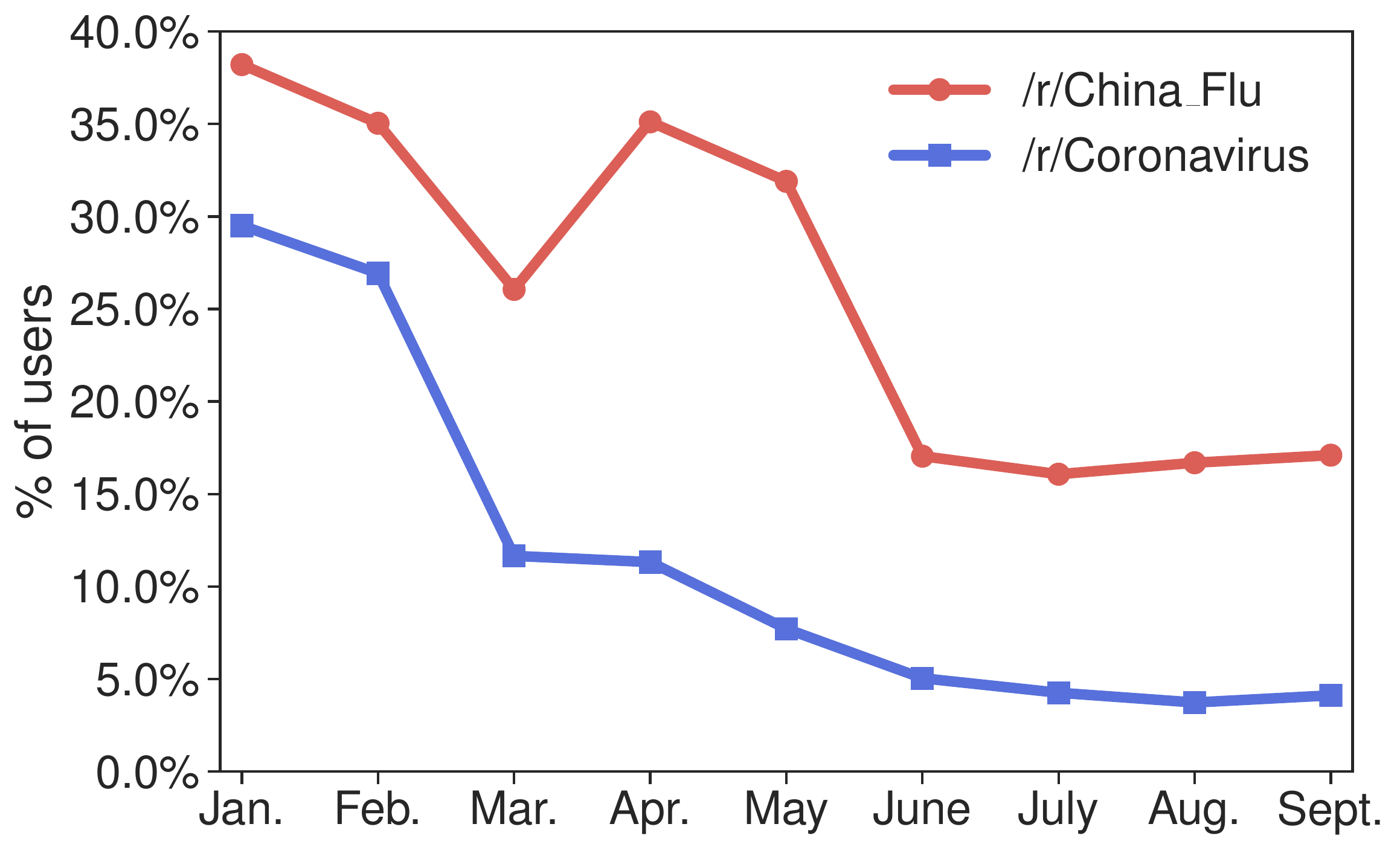}
    \caption{
    The proportion of users in \rchinaflu and \rcoronavirus whose comments contain 
    keywords ``china'' or ``chinese'' in that month.
    }
    \label{fig:china}
\end{figure}

\begin{table*}
\footnotesize
\begin{tabular}{ccp{15cm}}
\toprule
 &Score &                                                                                                                                                                        Comment \\
\midrule
  \multirow{11}{*}{ \rotatebox{90}{\rchinaflu} }
  &-104 & 1. Focus less on China and more on your US. You are on brink of collapse. \\
  &-103 &  2. You know what? You guys can talk about any government you would like but when you bring in a large group of people that is racism. That is not tolerated on this sub. Neither is Xenophobia, just because someone does something different from you does not make it wrong. \\
   &-97 & 3. Thought you guys claiming he was kidnapped by the big bad wolf CCP.  Turns out he was forced quarantined so he doesn't go out spreading the virus like a dumbass. Like what is happening in America. \\
   &-93 & 4. Is there a real problem? China preferred money and they're getting what they want. Donation is giving away something for free without any requirement or expectation of reciprocity.  \\
   &-91 & 5. Western countries got hit hard because the gov disregard Chinese government’s advice and refused to ask people to wear masks, not because they got news late. \\
\midrule
 \multirow{10}{*}{ \rotatebox{90}{\rcoronavirus} }
 &-467 & 1. the peak has passed, open the country \\
  &-452 &  2. The US response to the Corona Virus has been far better and more timely than most countries. The government will never look good no matter the response. Death never looks good. When the government did respond, they were attacked by the media for acting to harsh... \\
  &-431 & 3. The Coronavirus is very much under control in the USA. We are in contact with everyone and all relevant countries. CDC and World Health have been working hard and very smart. Stock Market starting to look very good to me! \\
  &-357 &  4. I mean it feels like 99\% of the world downplayed the virus. The see-I-told-ya-so’s can apply to lots of people. I don’t see the point of it now. \\
  &-324 &  5. A mix of medias have both downplayed the virus at some point. There is no use in making this into a political fight.. people make judgments on limited information and then shit hits the fan, it happens to everyone, there is no use in pointing fingers and wishing bad on anyone. \\
 \bottomrule
\end{tabular}
\caption{Most downvoted comments in \rchinaflu and \rcoronavirus.}
\label{tab:top_downvotes}
\end{table*}

The phrase ``China Flu'' is frequently used, especially by President Donald Trump, to blame 
China for the pandemic \cite{trumpchinaflu,joubin2020anti,schild2020go}.
Even though the founders of \rchinaflu claim that the community ``china flu'' was created 
before the virus was officially named, it may nevertheless attract people who are anti-China.
\figref{fig:china} compares the proportion of comments that contain keywords ``china'' or ``chinese''.
Indeed, \rchinaflu users talk more about China 
than \rcoronavirus users in all the months.
In January and February, when China was the epic center of coronavirus, 
users in both communities had a high rate of mentioning ``china'' or ``chinese''.
After that, when China contained the virus, \rchinaflu users still maintain a high interest in
discussing China-related topics, while users in \rcoronavirus seldom bring up China.
the blue line drops to a low level.

We also examine the most downvoted comments in these two communities.
By design, Reddit communities allow users to upvote or downvote the comments they read. 
The difference between the number of upvotes and downvotes a comment receives 
is referred to as ``score''.
The ranking system will display comments with the highest scores at the top of the page
and hide the ones with the lowest scores. 
The most downvoted comments can help us understand what opinions that the community
members dislike intensely.

\tableref{tab:top_downvotes} lists the top-5 most downvoted comments in \rchinaflu and \rcoronavirus.
All the comments in \rchinaflu are related to 
positive opinions
about China or the Chinese government. 
In contrast, downvoted comments in \rcoronavirus are mostly about downplaying the virus's seriousness.
These comments suggest that users in \rchinaflu pay much closer attention
to Chinese-related news and tend to show a negative opinion against China.

\section{New Users Choosing between \\\rchinaflu and \rcoronavirus}
\label{sec:traj}
To further delineate the 
declining overlap in active users of these two communities,
we investigate how {\em new} users chose to start at \rchinaflu or \rcoronavirus.
We group users based on which month they started at \rchinaflu
or \rcoronavirus and which community they participated in first. 
This allows us to examine different cohorts of users and zero in on the declining overlap of the active users each month.

\begin{figure}
    \center
    \includegraphics[width=0.35\textwidth]{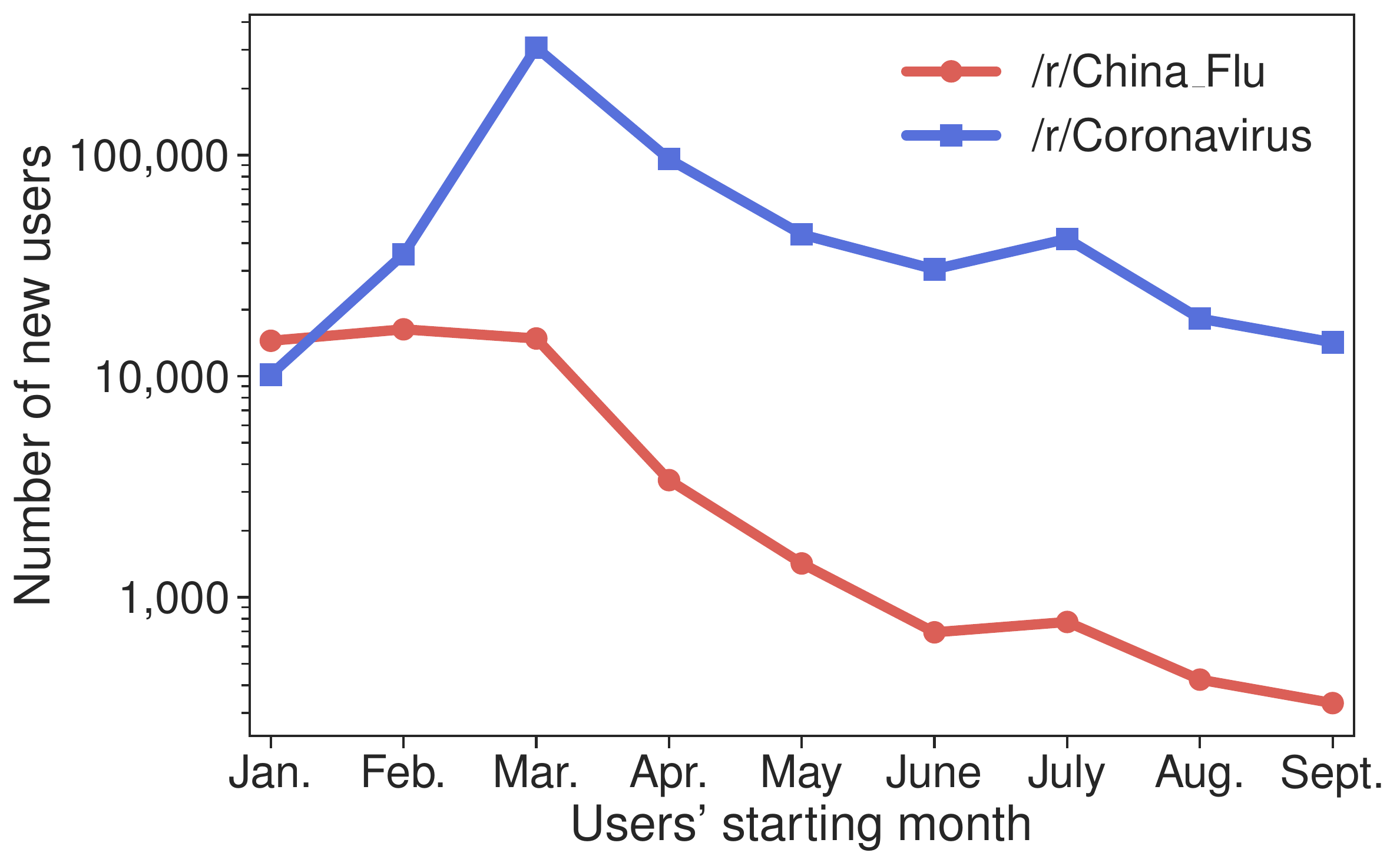}
    \caption{The number of users who start at \rchinaflu or \rcoronavirus per month in {\em log} scale. 
    Note that every user is only counted once here. 
    If the user commented in \rchinaflu first, they will not be counted as a new user in \rcoronavirus, and vice versa. }
    \label{fig:num_start}
\end{figure}

\figref{fig:num_start} shows the number of new users who started at \rchinaflu and \rcoronavirus each month
(Note that every user is only counted once. 
If a user commented in \rchinaflu first, they will not be counted as a new user in \rcoronavirus, and vice versa).
Since June, \rchinaflu has become much less active: fewer than 1000 new users %
joined this community
each month. In comparison, \rcoronavirus attracts new users at a high level.
In this section, we refer to users who chose to \textbf{\textit{start}} at \rchinaflu as {\em \rchinaflu users} and in \rcoronavirus as {\em \rcoronavirus users}.

We also collect users' commenting history on the entire Reddit platform from one month before they started at
\rchinaflu or \rcoronavirus to the end of September 2020. 
In January and February, all the new users in both communities are included.
From March to September, we include all the new users from \rchinaflu and randomly sample the same number 
of new users from \rcoronavirus, 
as there are many more new users each month in \rcoronavirus than in \rchinaflu.

\subsection{Language Usage Difference}

We characterize the behavioral differences between \rchinaflu users and \rcoronavirus users through the lens of language usage, 
as posting comments is the major activity for users on Reddit.
Similar to prior work \cite{tan2016lost,atkinson2019gets,althoff2016large}, 
we adopt the Jensen-Shannon Divergence~\cite{manning1999foundations} 
to measure the monthly unigram usage difference in comments as the language distance.
A larger distance indicates a larger difference in language usage.
For all text-related computations in this paper, we remove punctuation marks, urls,
and stopwords. We also stem the words during preprocessing. 
The preprocessing is implemented using Gensim \cite{rehurek_lrec}.
We also remove comments from moderators or bots 
to focus on community members' responses during this pandemic.

We measure the language distance between \rchinaflu users and \rcoronavirus users in two contexts.
First, we measure their language distance 
before joining these two communities to
characterize their differences before COVID-19.
To do that, we aggregate the entirety of each user's comments on the Reddit platform 
one month before starting in \rchinaflu or \rcoronavirus  (all the communities).
Second,
we aggregate each user's comments in \rchinaflu and \rcoronavirus to measure the language distance in these two communities.

\begin{figure}
    \center
    \includegraphics[width=0.35\textwidth]{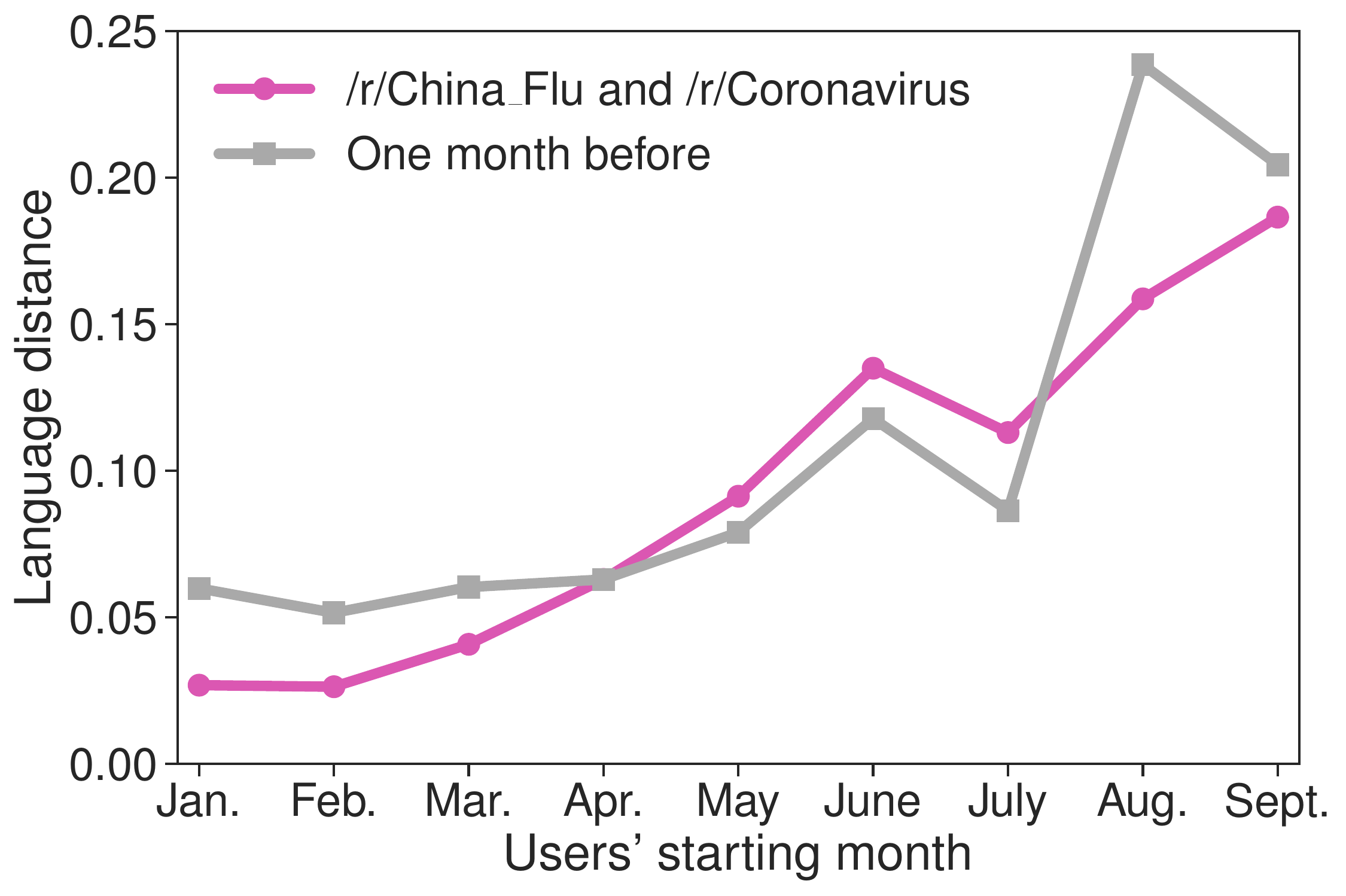}
    \caption{The language distance between \rchinaflu and \rcoronavirus users who started in the same month.
    The gray line measures the language distance of users' comments on the entire Reddit platform one month before
    joining \rchinaflu or \rcoronavirus.
    The pink line measures the language distance of users' entire comments (from the starting month to September)
    in \rchinaflu or \rcoronavirus.
    }
    \label{fig:language_distance}
\end{figure}

\figref{fig:language_distance} summarizes the monthly language distance between users who started
at \rchinaflu or \rcoronavirus, as measured by the Jensen-Shannon Divergence.
The language used by new users who started at \rchinaflu and \rcoronavirus becomes more and more different, suggesting a widening separation between these two communities.
Moreover, the language distance between \rchinaflu users and \rcoronavirus users one month before joining these two communities also increases.
This observation indicates that these two communities attract users who were already different 
before joining COVID-related discussions.
A closer look shows that the language distances of these two measures were low in January and February, especially in \rchinaflu and \rcoronavirus.
This suggests that the discussions at these two communities were similar, and users who joined these two communities at that time were also similar.
Peak similarity appeared in February when the official 
announcement was made (\figref{fig:feb17}). 
The strict moderating rules implemented in \rcoronavirus and allowing more relaxed discussions in \rchinaflu
may have played roles in this shift.
We further note a dip in July for both lines. 
One explanation could be due to the second wave of infected cases in the U.S. and other countries, 
more new users are looking for COVID-19 communities to follow, and they tend to be more similar
(We also observe a peak of activity in July in \figref{fig:total_num}).

\begin{figure}
    \center
    \includegraphics[width=0.35\textwidth]{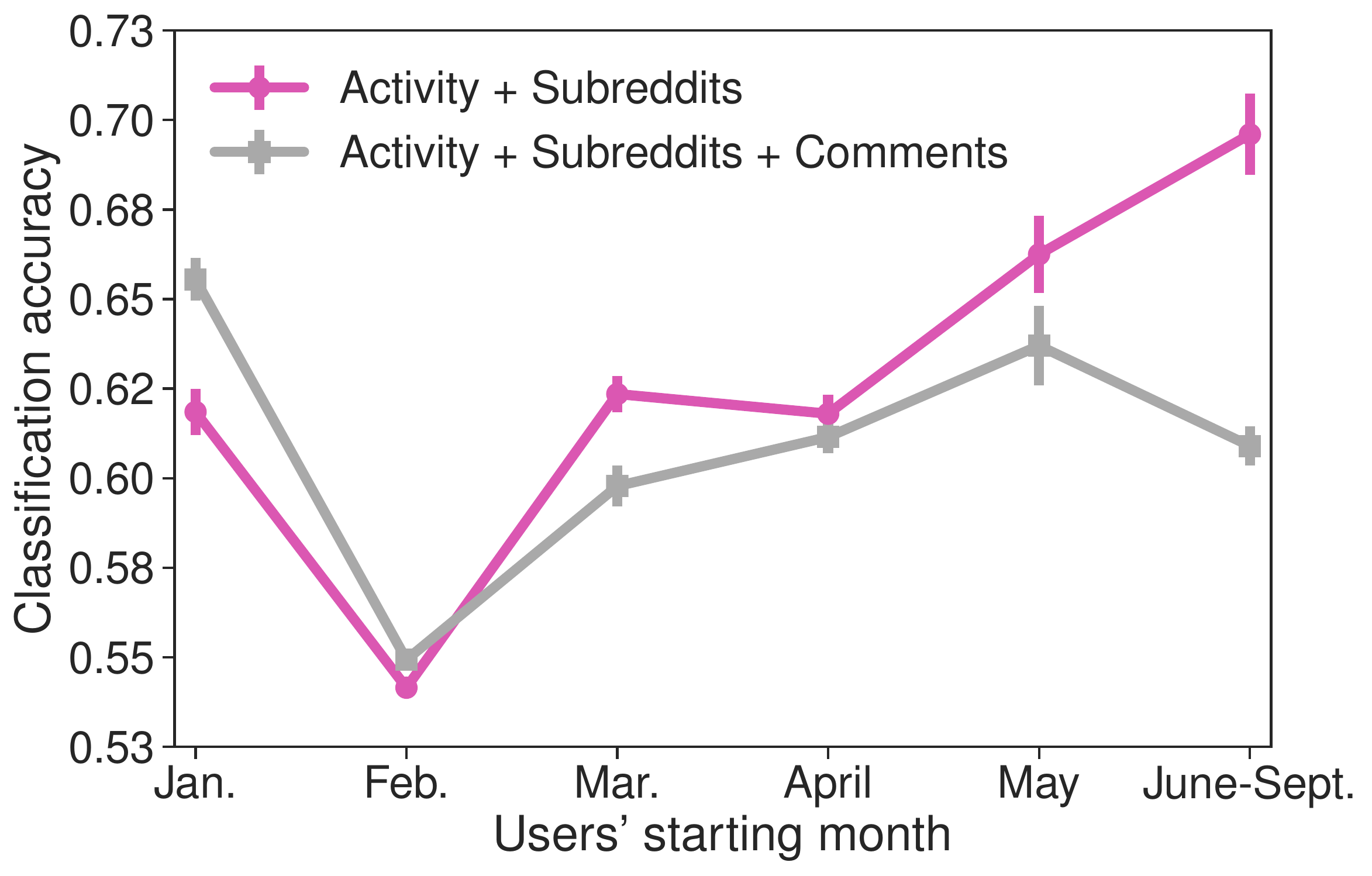}
    \caption{
    Monthly accuracy of determining which community (\rchinaflu or \rcoronavirus) 
    a new user will start at.
    The pink line shows the monthly accuracy using Activity and Subreddits features, while the gray line
    shows the accuracy using Activity, Subreddits, and Comments features. 
    }
    \label{fig:accuracy}
\end{figure}

\subsection{In Which Community will a User Start?}
\label{sec:pred_start}

Another way to understand the differences between users who chose to start at \rchinaflu or \rcoronavirus 
is to measure their predictability. 
We formulate a prediction task: 
given the users' activity on the Reddit platform one month before starting at \rchinaflu or \rcoronavirus,
can we predict which community they will choose?
For each month, we randomly sample 1000 users who started at \rchinaflu or \rcoronavirus, respectively.
As there are not enough users from June to September in \rchinaflu, we combine users in these months and randomly
sample 1000 users from each community.
This task is thus a balanced prediction task, where the majority baseline accuracy is 50\%.
We extract the following features from users' comments one month before joining \rchinaflu or \rcoronavirus:

\begin{itemize}
  \item Activity: these features include the number of comments made by the user, average comment length, 
    and the entropy of all subreddits where the user has been active.
  \item Subreddits: the bag of subreddits (similar to the bag of words (BOW)), 
  based on the all subreddits where the user has been active one month 
  before joining \rchinaflu or \rcoronavirus. 
  We remove subreddits that appeared less than five times in the training data.
  \item Comments: the bag-of-words (BOW) from users' comments in all the communities, 
   one month before joining \rchinaflu or \rcoronavirus.
\end{itemize}

\begin{table*}[t]
\scriptsize
\begin{tabular}{l|cccccc}
\toprule
&                          Jan. &                                Feb. &                           Mar. &                                Apr. &                             May. &                        June-Sept. \\
\midrule
 \multirow{10}{*}{ \rotatebox{90}{\rchinaflu} }  &        /r/collapse (-0.4) &       /r/singapore (-0.6) &       /r/wuhan\_flu (-0.4) &      /r/conspiracy (-0.4) &           /r/china (-0.4) &     /r/conspiracy (-0.2) \\
  &       /r/worldnews (-0.4) &           /r/china (-0.5) &      /r/conspiracy (-0.4) &     /r/cringetopia (-0.3) &      /r/conspiracy (-0.2) &     /r/wuhan\_flu (-0.2) \\
  &       /r/singapore (-0.3) &        /r/collapse (-0.4) &        /r/preppers (-0.4) &       /r/wuhan\_flu (-0.3) &  /r/politicalcompassmemes &   /r/conservative (-0.2) \\
  &        /r/preppers (-0.3) &          /r/taiwan (-0.3) &        /r/collapse (-0.3) &           /r/china (-0.3) &            /r/gifs (-0.1) &       /r/covid19positive \\
  &            /r/news (-0.3) &  /r/wallstreetbets (-0.3) &  /r/wallstreetbets (-0.3) &   /r/worldpolitics (-0.2) &        /r/animalsonreddit &          /r/china (-0.2) \\
  &           /r/china (-0.3) &          /r/canada (-0.3) &       /r/singapore (-0.3) &  /r/politicalcompassmemes &       /r/whatcouldgowrong &   /r/thelastofus2 (-0.2) \\
  &   /r/hongkong (-0.3) &        /r/preppers (-0.3) &   /r/entertainment (-0.3) &    /r/historymemes (-0.2) &        /r/libertarianmeme &  /r/okbuddyretard (-0.2) \\
  &  /r/wallstreetbets (-0.3) &           /r/italy (-0.2) &      /r/the\_donald (-0.3) &        /r/hongkong (-0.2) &         /r/gadgets (-0.1) &            /r/art (-0.2) \\
  &             /r/nba (-0.2) &  /r/presidentialracememes &           /r/korea (-0.3) &          /r/europe (-0.2) &           /r/diwhy (-0.1) &     /r/mildlyinteresting \\
  &          /r/canada (-0.2) &      /r/bestoflegaladvice &           /r/china (-0.3) &   /r/iamatotalpieceofshit &       /r/wuhan\_flu (-0.1) &  /r/anime\_titties (-0.1) \\
\midrule
\multirow{10}{*}{ \rotatebox{90}{\rcoronavirus} }   &       /r/roastme (0.3) &      /r/breadtube (0.3) &            /r/nfl (0.3) &  /r/insaneparents (0.3) &       /r/askreddit (0.3) &        /r/politics (0.5) \\
  &    /r/the\_donald (0.2) &  /r/latestagecapitalism &         /r/hockey (0.3) &      /r/askreddit (0.3) &        /r/politics (0.3) &             /r/nba (0.3) \\
  &   /r/cringetopia (0.2) &        /r/epstein (0.2) &   /r/rocketleague (0.3) &       /r/politics (0.3) &            /r/news (0.3) &       /r/askreddit (0.3) \\
  &         /r/mgtow (0.2) &       /r/patriots (0.2) &     /r/indieheads (0.3) &     /r/technology (0.2) &        /r/facepalm (0.3) &  /r/politicalhumor (0.2) \\
  &  /r/natureisfuckinglit &    /r/90dayfiance (0.2) &      /r/parenting (0.3) &            /r/nba (0.2) &    /r/whitepeopletwitter &       /r/winstupidprizes \\
  &  /r/rickandmorty (0.2) &            /r/cfb (0.2) &            /r/nba (0.3) &  /r/amitheasshole (0.2) &             /r/nfl (0.2) &            /r/news (0.2) \\
  &       /r/science (0.2) &   /r/reactiongifs (0.2) &         /r/soccer (0.2) &  /r/sandersforpresident &         /r/xboxone (0.2) &       /r/oddlysatisfying \\
  &    /r/bad\_cop\_no\_donut &   /r/bodybuilding (0.2) &  /r/modernwarfare (0.2) &      /r/humansbeingbros &  /r/animalcrossing (0.2) &             /r/aww (0.2) \\
  &     /r/pan\_media (0.2) &      /r/choosingbeggars &            /r/wtf (0.2) &         /r/boston (0.2) &     /r/therewasanattempt &       /r/humansbeingbros \\
  &          /r/meme (0.2) &        /r/woooosh (0.2) &  /r/bikinibottomtwitter &        /r/science (0.2) &   /r/dundermifflin (0.1) &  /r/nintendoswitch (0.2) \\
  \bottomrule
\end{tabular}
\caption{The monthly top-10 subreddits with the highest coefficients for predicting users' starting community 
(\rchinaflu or \rcoronavirus). Due to space limit, we hide the coefficients of subreddits whose names are long.}
\label{tab:monthly_subs_coefs}
\end{table*}

To assess the prediction performance, we measure accuracy using stratified 
five-fold nested cross-validation with a standard $\ell_2$-regularized logistic regression classifier. 
All the features are scaled to [0, 1] based on the training data.

\figref{fig:accuracy} presents the monthly accuracy of predicting
which community (\rchinaflu or \rcoronavirus) a new user is going to start at based on her activity
on the Reddit platform one-month before.
First, the accuracy for each month is much higher than 50\%. 
It shows that users' previous activity history (also known as genealogy \cite{tan2018tracing}) matters, 
and using historical activities can outperform the random baseline
by a significant margin. 
Second, we observe a similar trend as the language distance:
The accuracy hits bottom in February, indicating that the users who joined 
\rchinaflu or \rcoronavirus are hard to distinguish in that month.
After that month, the accuracy goes up, indicating an increment of distinction.
Interestingly, the best accuracies for months after March are achieved
by only using Activity 
and Subreddits features. 
Adding features from comments in the model is not helpful. 
It is likely due to the sparsity of words in our relatively small dataset for the prediction task,
as the comments can come from many different communities.
This observation further shows that genealogical relations may provide more robust signals than textual content.

\para{Subreddit importance analysis.}
\figref{fig:accuracy} shows that the best prediction accuracy is achieved by using Activity 
and Subreddit features. 
A close examination of the coefficients can reveal the subreddits that are strong indicators.
Here we rank subreddits month by month based on their coefficients in the prediction model,
and the results are summarized in \tableref{tab:monthly_subs_coefs}.
First, \communityname{/r/china} is in the top-10 subreddits for \rchinaflu across all the months. 
This is a community for discussing China-related topics, and most of its posts and comments are anti-China.\footnote{\url{https://www.reddit.com/r/china/}.}
It indicates that people who dislike China or China-related news are more likely to 
join \rchinaflu. 
Subreddits that focus on places near China geographically are also strong indicators 
(e.g., \communityname{/r/singapore}, \communityname{/r/hongkong}, and \communityname{/r/taiwan}). 
Second, some well-known extreme subreddits, such as \communityname{/r/conspiracy} and \communityname{/r/wuhan\_flu},
have been strong predictors for \rchinaflu users since March. 
Interestingly, we do not see them in January or February. 
We also do not see them in subreddits that predict \rcoronavirus users.

\section{User Movement between \\ \rchinaflu and \rcoronavirus}
\label{sec:movement}

\begin{figure*}
    \center
    \begin{subfigure}[t]{0.4\textwidth}
        \includegraphics[width=0.9\textwidth]{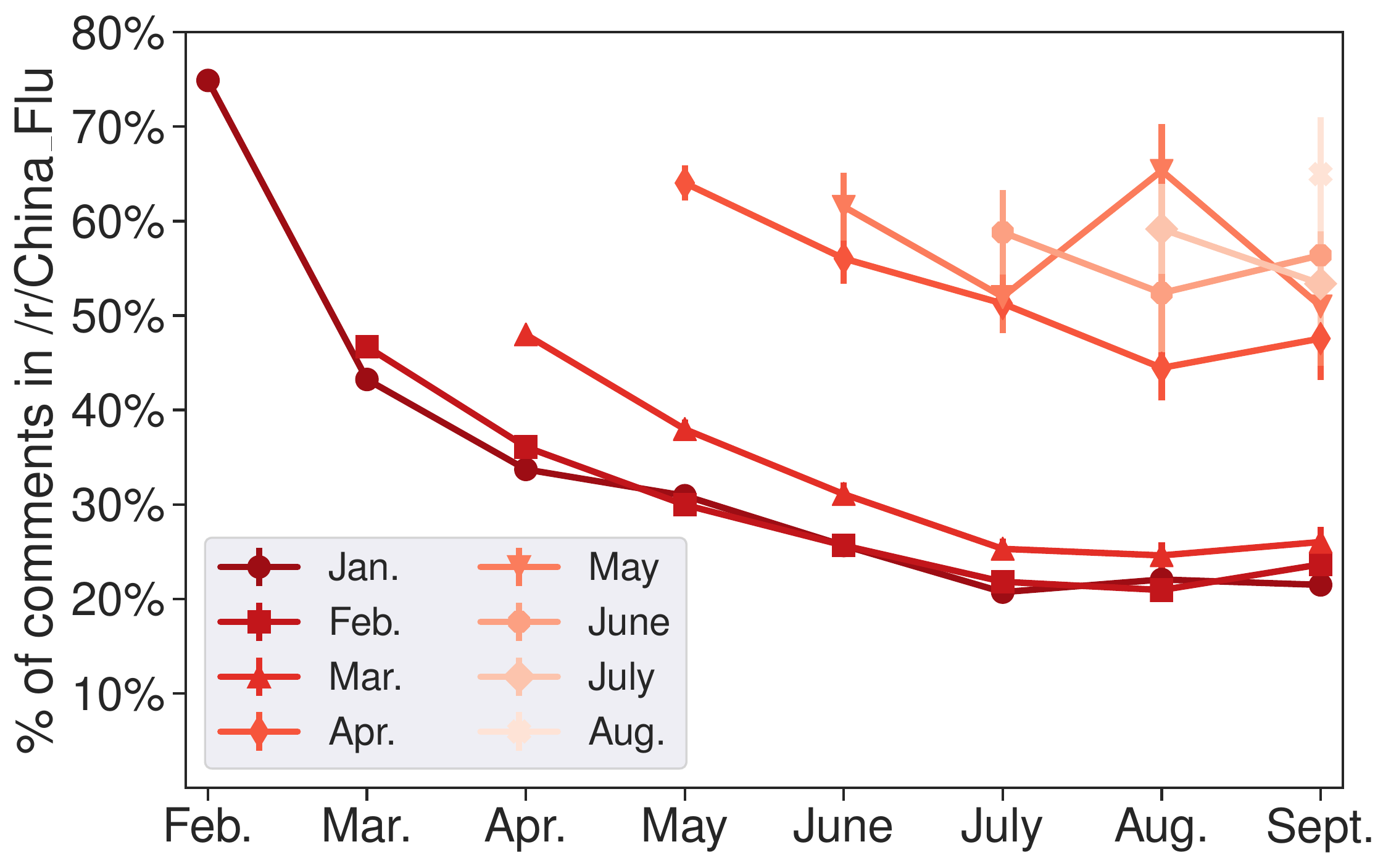}
        \caption{\rchinaflu users' monthly activity.}
        \label{fig:start_china_prop}
    \end{subfigure}
    \begin{subfigure}[t]{0.4\textwidth}
        \includegraphics[width=0.9\textwidth]{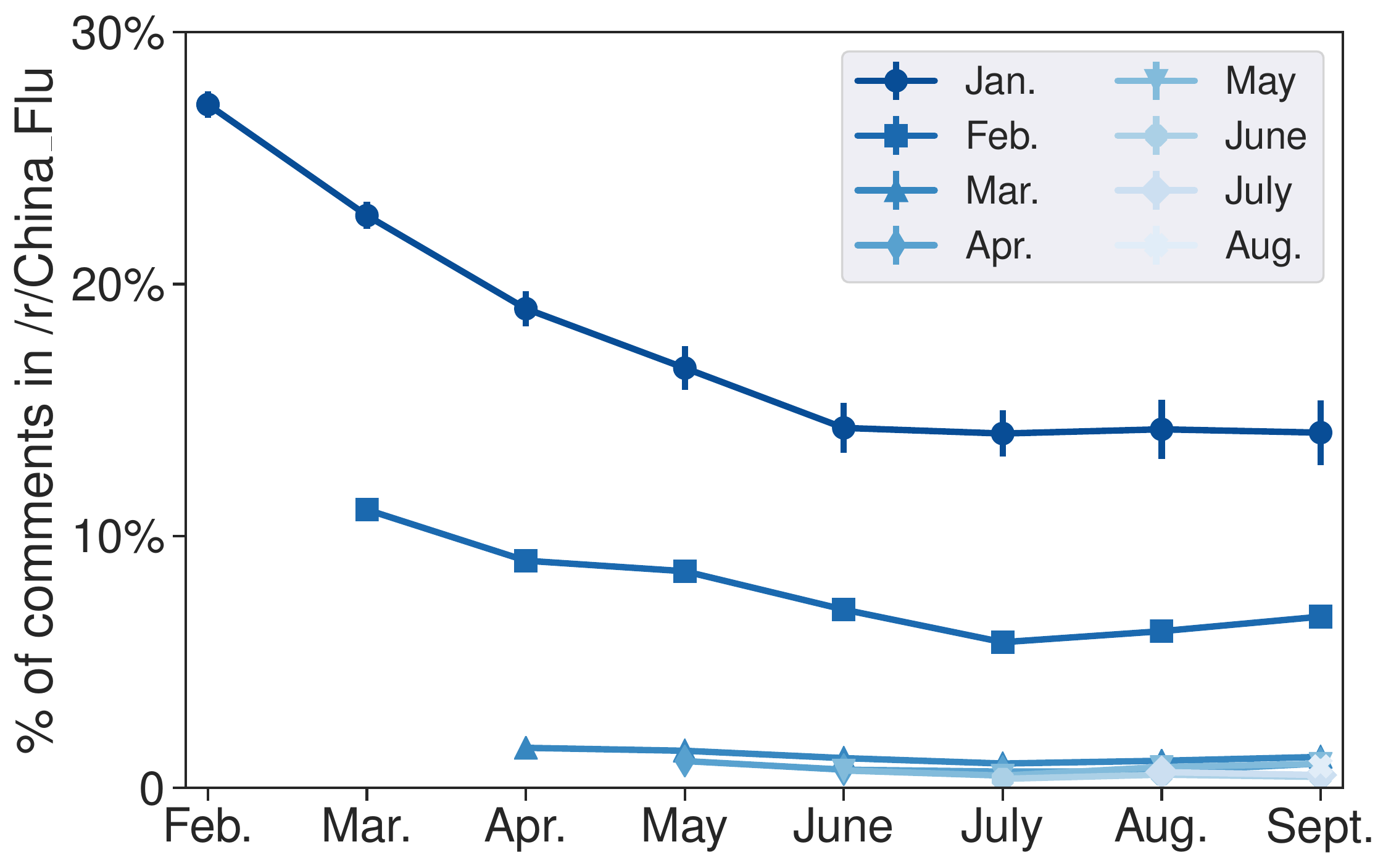}
        \caption{\rcoronavirus users' monthly activity.}
        \label{fig:start_corona_prop}
    \end{subfigure}
    \caption{
     The monthly proportion of activity in \rchinaflu for users who started in different months in 
     \rchinaflu (\figref{fig:start_china_prop}) and \rcoronavirus (\figref{fig:start_corona_prop}), respectively.
     Each line represents users who started in a month from January to August.
    }
    \label{fig:start_prop}
\end{figure*}

We have identified clear signals of whether a new user chooses to start in \rchinaflu or \rcoronavirus in their genealogy, but it remains unclear how stable this initial membership is.
Hence, we investigate user movement between these two
communities.

\subsection{How do Users Split their Activity in \rchinaflu and \rcoronavirus?}

We first analyze how users split their activity in \rchinaflu and \rcoronavirus, and how it differs over time.
To illustrate this, we show the monthly activity split for users who started in different months in \rchinaflu or \rcoronavirus, respectively.
Each curve in \figref{fig:start_prop} is obtained by aggregating users who started in the same month in \rchinaflu or \rcoronavirus and then
calculating their proportion of activity in \rchinaflu among these two communities over time.

\rchinaflu users who started in January (the darkest red) have a sharp drop 
in the following months, from 75\% in February to below 30\% since June (\figref{fig:start_china_prop}). 
This indicates that the majority of these users' activity has shifted from \rchinaflu to \rcoronavirus during this period.
We observe a similar trend for \rchinaflu users who started in February and March.
However, there is a clear separation between \rchinaflu users who started 
in the first three months and the remaining months. 
\rchinaflu users who started after March tend to maintain a high level of activity in \rchinaflu (above 50\%) over time.

In comparison (\figref{fig:start_corona_prop}), \rcoronavirus users who started in January and February 
sustained a certain level of activity in \rchinaflu (around 15\% for users who started in January and around 8\% for users who started in February).
However, users who joined later have little activity in \rchinaflu. 
These comparisons echo the clear separations between the two communities we observed previously, especially from April to September.

\begin{figure*}
    \center
    \begin{subfigure}[t]{0.32\textwidth}
        \includegraphics[width=0.95\textwidth]{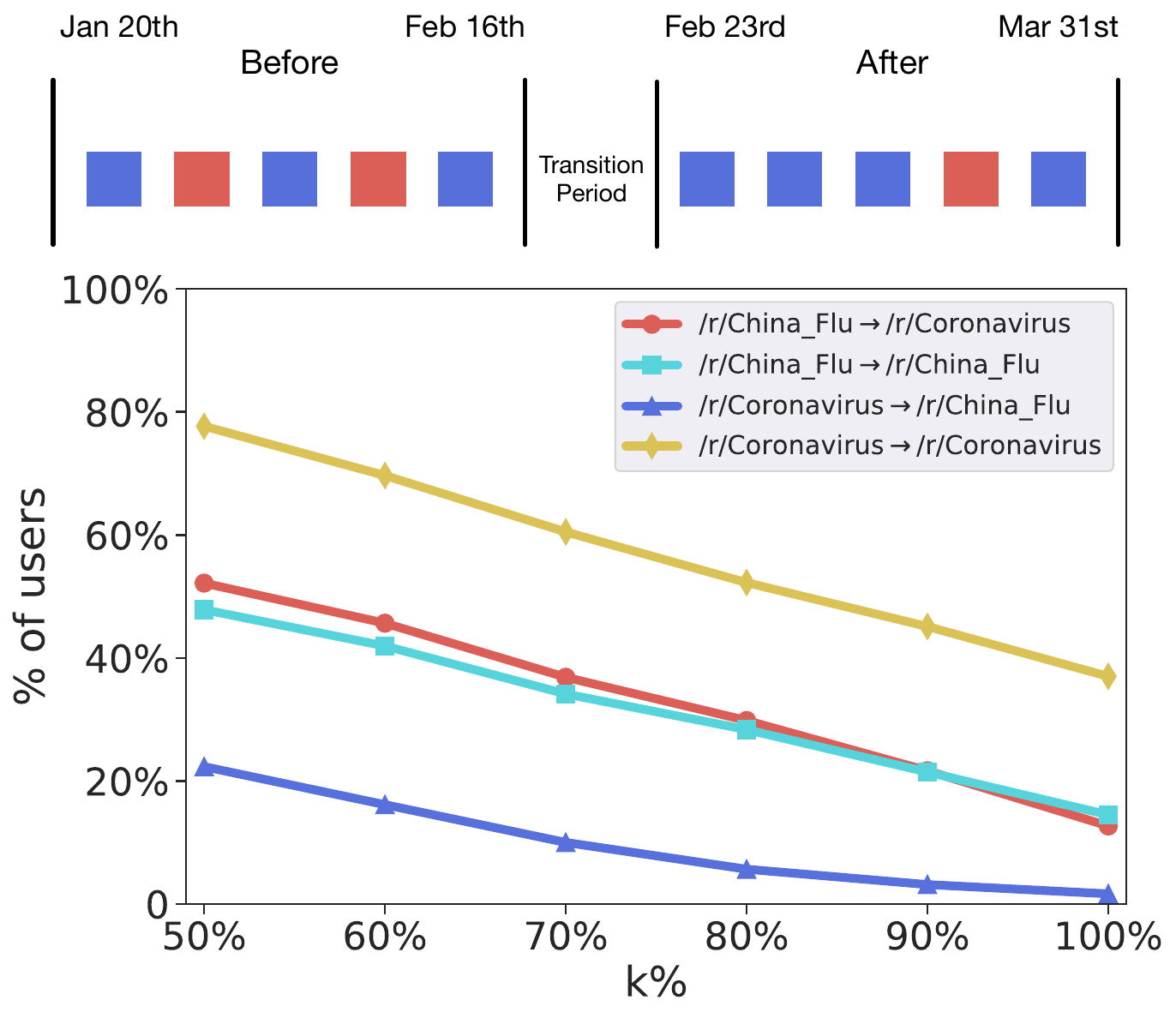}
        \caption{Feb. 16th to Feb. 23rd as the transition window for \rchinaflu and \rcoronavirus.}
        \label{fig:movement_one}
    \end{subfigure}
    \hfill
    \begin{subfigure}[t]{0.32\textwidth}
        \includegraphics[width=0.95\textwidth]{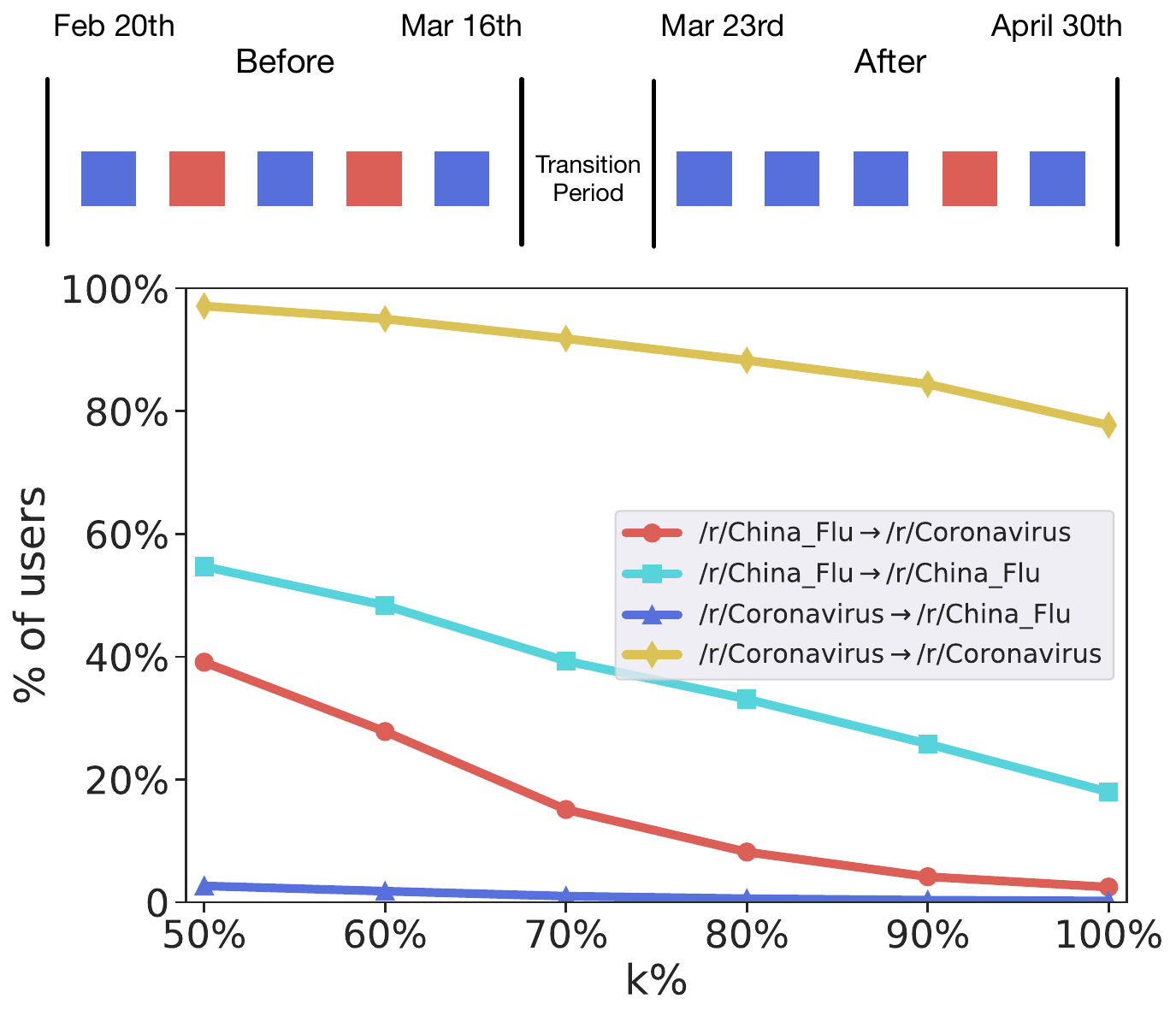}
        \caption{Mar. 16th to Mar. 23rd as the transition window for \rchinaflu and \rcoronavirus.}
        \label{fig:movement_two}
    \end{subfigure}
    \hfill
    \begin{subfigure}[t]{0.32\textwidth}
        \includegraphics[width=0.95\textwidth]{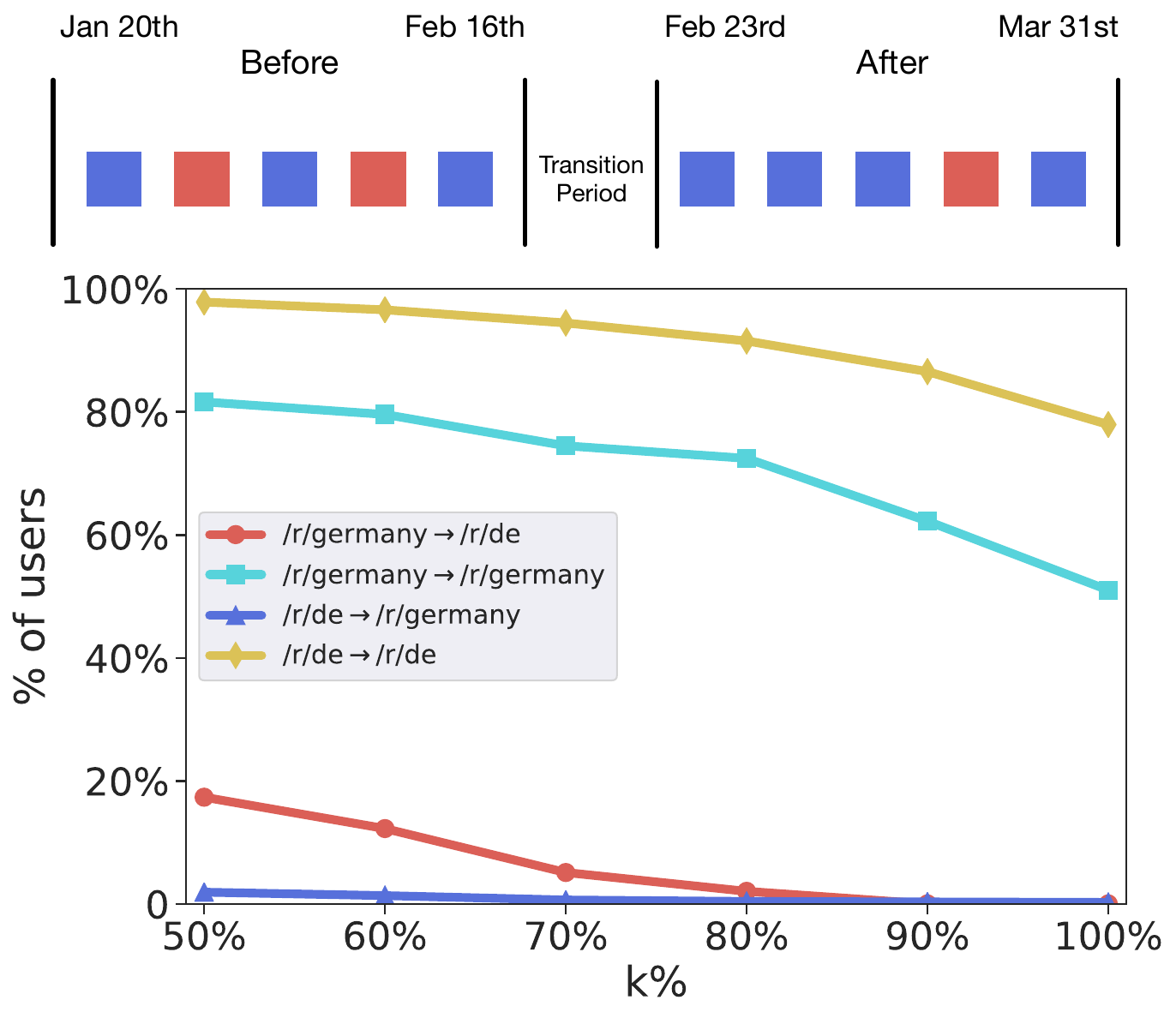}
        \caption{Feb. 16th to Feb. 23rd as the transition window for \communityname{/r/germany} and 
        \communityname{/r/de}.}
        \label{fig:movement_de}
    \end{subfigure}
    \caption{
     \figref{fig:movement_one} shows the proportion of users who moved between \rchinaflu and \rcoronavirus 
     before and after the transition window (from February 16th to February 23rd). 
     \figref{fig:movement_two} applies this movement framework to the transition window one month later 
     (from March 16th to March 23rd). 
     For further comparison, \figref{fig:movement_de} applies this movement framework to another two highly-related communities, 
     \communityname{/r/germany} and \communityname{/r/de} (transition window from February 16th to February 23rd).
     The blue and red squares are only for illustrative purposes.
     They represent a user's activity in \rchinaflu and \rcoronavirus, respectively.
    }
    \label{fig:movement}
\end{figure*}

\subsection{A Movement Analysis Framework}
Given that early \rchinaflu users commented less in \rchinaflu and more in \rcoronavirus over time, 
we further investigate users who decided to move from \rchinaflu to \rcoronavirus and the potential factors that may influence their choices. 
To understand this behavior, an important question is to determine users' memberships in these two communities.
One way is to define users who only commented in \rchinaflu as \rchinaflu members and 
apply the same rule for detecting \rcoronavirus members (note that the definition differs from the previous section, which bases on a user's starting community. We thus use ``members'' to signify the difference).
However, as shown in \figref{fig:common}, a large proportion of \rchinaflu users also commented
in \rcoronavirus, especially in the early months (February to April).  
To provide a more complete picture, we develop a movement analysis framework and 
use an activity ratio threshold $k\%$ to determine a user's membership
in these two communities.

A user is a member of \rchinaflu in a time window if more than 
$k\%$ of their comments {\em in these two communities} are posted in \rchinaflu. 
The same rule is used for determining \rcoronavirus members.
For example, when $k\%=60\%$, users who made more than 60\% of their comments in \rchinaflu are \rchinaflu members.
Users who made more than 60\% of their comments in \rcoronavirus are \rcoronavirus members.
Users who do not fall into these two categories are discarded, as their memberships are deemed uncertain.
In particular, when $k\%=100\%$, users who only commented in \rchinaflu are \rchinaflu members, and 
users who only commented in \rcoronavirus are \rcoronavirus members.

To examine users who moved from one community to the other, we need to set a transition window and compare 
their memberships before and after that transition window. 
As discussed in \figref{fig:feb17}, on February 17th, the Reddit platform made \rcoronavirus the
official community, and more relaxed discussions are allowed in \rchinaflu. 
This decision may have motivated users to move between these two communities. 
Thus, we set the week of February 17th as a transition window (February 16th to February 22nd) 
and measure users' memberships before and after this transition window,
illustrated in \figref{fig:movement_one}. 
The users' commenting activity before February 16th is used for determining their memberships before the transition window,
and their activity after February 22nd but before March 31st 
is used for determining memberships after the transition window.
We also require users to have at least five comments in these two communities before the transition and an additional five comments
after the transition to be included in this study.
There are in total 11,238 such users.
If a user's memberships before and after the transition are different, 
then it means that she moved from one community to the other.

\figref{fig:movement_one} shows the movement analysis results.
With varying $k\%$, the number of \rchinaflu users who moved to \rcoronavirus (red line) during
the transition window is similar to the number of users who stayed (cyan line).
It shows that around 50\% of \rchinaflu members have 
moved to \rcoronavirus during this transition window, robust to the definition of membership ($k\%$).
Specifically, 
13\% of users made a dramatic shift, from only commenting in \rchinaflu to only commenting in \rcoronavirus ($k\%=100\%$).
The ratio of users who left is always around 50\% regardless of $k$.
In comparison, a small proportion of \rcoronavirus members moved to \rchinaflu (blue line). 
Most of them stayed in \rcoronavirus (yellow line).
When we choose $k=100$, the number of users who went from 100\% \rcoronavirus to 100\% \rchinaflu is close to zero.

To further examine such movement's robustness,
we defer the transition window a month later and conduct the same analysis 
to find users who were moving between \rchinaflu and \rcoronavirus in March (\figref{fig:movement_two}).
The results demonstrate that few users have moved between these two communities. 
We also attempt to shift the transition window to the following months. 
The analyses show that the proportion of users who moved 
between these two communities in those months are even lower than in March.

In addition, we apply this movement analysis framework to another pair of highly-related communities, \communityname{/r/germany}
and \communityname{/r/de} (\figref{fig:movement_de}). 
Both of these two communities are for discussions about Germany.
The primary language in \communityname{/r/germany} is English, while 
the primary language in \communityname{/r/de} is German.
When we set the transition window from February 16th to February 23rd, very few users moved 
from one to the other (similar to \figref{fig:movement_two}). 
This trend is consistent for \communityname{/r/germany}
and \communityname{/r/de} when we defer the transition window to later months.

In summary, the movement analysis results in \figref{fig:movement} suggest that there is a significant
number of \rchinaflu users moved to \rcoronavirus in February. 
This movement may connect with Reddit's decision to make \rcoronavirus the official community for COVID-19 and
allow more relaxed discussion in \rchinaflu. We do not see movement at this frequency in the following months
between these two communities or other highly-related communities.
Moreover, compared with \rchinaflu users, many fewer \rcoronavirus users moved to \rchinaflu in February.

\subsection{Predicting the Movement from \rchinaflu to \rcoronavirus}

\rchinaflu is in flux in February: about half of \rchinaflu members moved to \rcoronavirus.
We formulate another prediction task to compare users who chose to stay or leave:
given the members of \rchinaflu and their activity prior to the transition window in February, can we predict who 
will stay or leave?
Here we define users who were \rchinaflu members before and after the transition window as ``Stay'' users,
and users who moved from \rchinaflu to \rcoronavirus as ``Leave'' users.
To ensure a balanced prediction task, for each threshold $k\%$, we randomly sample the same number of 
Stay users from Leave users (except for $k\%=100\%$ where we do it reversely as there are more Stay users than Leave users). This is thus a balanced prediction task, and the random baseline is 50\%.
The task allows us to understand the dynamic membership in \rchinaflu and 
its formation in the context of highly-related communities.

For this prediction task, we extract the following features from users' comments before the transition window 
(from January 20th to February 15th):

\begin{itemize}[leftmargin=*,topsep=0pt]
  \item Activity: these features include the number of comments made by the user, average comment length,
  mention of ``china'' or ``chinese'' in \rchinaflu. 
    We also add the entropy of all subreddits where the user has been active. 
  \item Subreddits: the bag-of-subreddits (similar to bag of words) based on all subreddits where the user 
            has been active. \rchinaflu and \rcoronavirus are excluded.
  \item Comments: the bag-of-words (BOW) from aggregated comments in all the communities, excluding 
  \rchinaflu and \rcoronavirus.
  \item \rchinaflu Comments: the bag-of-words (BOW) from aggregated comments in \rchinaflu.
\end{itemize}

For each $k\%$ threshold, we measure accuracy using stratified five-fold nested cross-validation 
with a standard $\ell_2$-regularized logistic regression classifier, which is the same one we used in 
the previous section.
All the features are scaled to [0, 1] based on the training data.

\begin{figure}
    \center
    \includegraphics[width=0.35\textwidth]{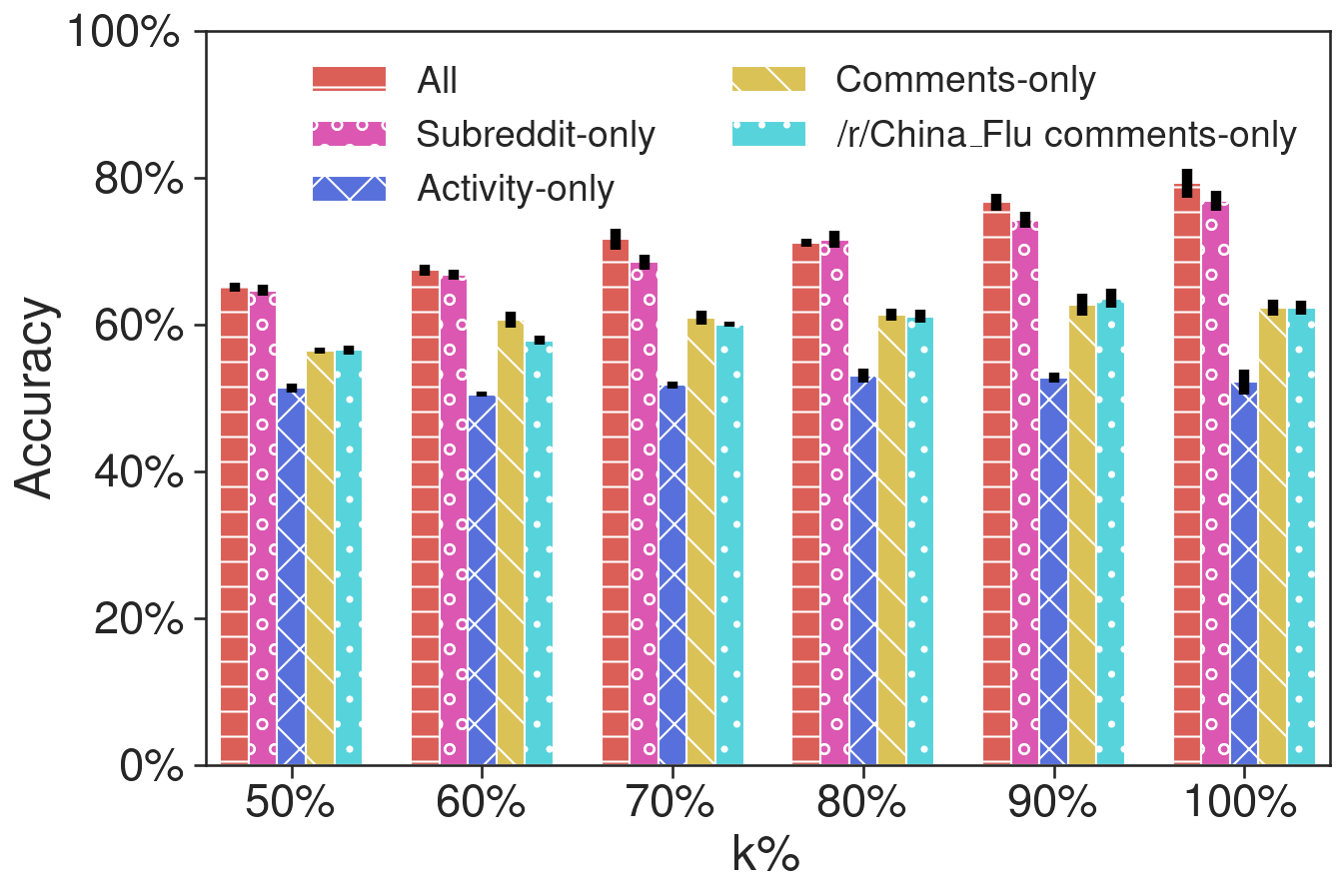}
    \caption{The performance of predicting users who would leave \rchinaflu for \rcoronavirus during the transition window
    (February 16th to 23rd) with different threshold $k\%$.}
    \label{fig:movement_accu}
\end{figure}

The prediction results are summarized in \figref{fig:movement_accu}. 
First, with a growing $k\%$, it is easier to predict users who will stay or leave.
For example, when $k\% = 100\%$, the prediction accuracy is as high as 80\%.
We see a near monotonic trend between $k\%$ and the prediction accuracy for each feature set. 
It suggests that for highly active users in \rchinaflu, 
the users who chose to stay or leave are substantially different and easy to distinguish. 
Second, the subreddits at which users have also been active are the strongest indicators of users' movement.
Only using the subreddits as the feature (Subreddit-only) 
can achieve comparable accuracies to using all the features (All), and
outperforms other types of features by a significant margin.

Inspired by this observation, we analyze the coefficients of subreddits in the model. 
\tableref{tab:movement_subs_coefs} summarizes the top-10 subreddits with the highest coefficients for 
predicting Stay and Leave users when choosing $k\% = 100\%$.
First, the subreddit with the highest coefficient for Stay users is \communityname{/r/the\_donald}.
This community was dedicated to supporting the U.S. President Donald Trump and was banned in March 2020 
due to racist content and hateful speech~\cite{bandonald}.
It suggests that a large proportion of Stay users are Donald Trump supporters. 
In contrast, \communityname{/r/sandersforpresident}, the community for Bernie Sanders
has the highest coefficient for predicting Leave users, 
indicating that many Leave users are Bernie Sanders supporters.
Besides, \communityname{/r/conspiracy}, \communityname{/r/preppers}, and \communityname{/r/collapse} 
are strong signs for users to stay in \rchinaflu. This is also reflected in founders' activity 
(\tableref{tab:parents}).  
In contrast, some science-oriented subreddits, such as \communityname{/r/askscience}, 
\communityname{/r/dataisbeautiful}, and \communityname{/r/covid19}, appear on the table's right column.
It indicates that science enthusiasts were leaving \rchinaflu for \rcoronavirus.

\begin{table}
\small
\begin{tabular}{lc|lc}
\toprule
Predicting Stay &  Coefs &   Predicting Leave &  Coefs \\
\midrule
        /r/the\_donald &   -1.4 &  /r/sandersforpresident &          1.1 \\
        /r/conspiracy &   -1.4 &       /r/dataisbeautiful &          1.0 \\
          /r/preppers &   -1.3 &               /r/covid19 &          1.0 \\
             /r/china &   -0.9 &                 /r/books &          0.9 \\
     /r/wellthatsucks &   -0.9 &            /r/askscience &          0.9 \\
          /r/collapse &   -0.7 &          /r/idiotsincars &          0.8 \\
          /r/facepalm &   -0.7 &                   /r/nba &          0.8 \\
    /r/showerthoughts &   -0.7 &       /r/leagueoflegends &          0.7 \\
             /r/funny &   -0.6 &            /r/coolguides &          0.7 \\
 /r/explainlikeimfive &   -0.6 &         /r/worldpolitics &          0.6 \\
\bottomrule
\end{tabular}
\caption{The top-10 subreddits with the highest coefficients for predicting Stay or Leave users ($k\% = 100\%$).}
\label{tab:movement_subs_coefs}
\end{table}

\section{Concluding Discussion}
\label{sec:discussion}

In this work, we study the emergence of two highly-related communities on Reddit, \rchinaflu and \rcoronavirus,
during the COVID-19 pandemic.
We take a user-centered perspective and characterize user trajectories in these two communities
from January to the end of September. 
We show that users who joined these two communities were similar in the first three months.
After that, as the pandemic continues to unfold, their differences steadily increase.
We further examine the user movement between these two communities.
We find that users who started at \rchinaflu from January to March reduced their activity at \rchinaflu later,
while users who started in the following months remained highly ``loyal''.
A newly designed movement analysis framework reveals that around 50\% of \rchinaflu members
moved to \rcoronavirus in February, when \rcoronavirus became the official COVID-19 community on Reddit.
This movement turns out to be highly predictable based on other subreddits users were formerly active in.

\para{Limitations.}
The findings in our work are subject to several limitations. 
First, the causal reasons for users to move from \rchinaflu to \rcoronavirus are not established.
Potential reasons include 1) making \rcoronavirus the official community on the platform; 
2) the strict moderation rules implemented in \rcoronavirus, which may also attract users who seek serious discussions about COVID-19;
3) The anti-china sentiment of ``China\_Flu'', which may drive people who have complementary views about China to leave.
Our study is limited to revealing correlations.
Second, even though our findings reveal many similarities between these two communities at the beginning stage,
there are potentially other ways in which initial members of two communities differ that are not easy to capture
from our data.
Surveys or interviews with early members or moderators may help us better understand this development.
Moreover, most of the data has been collected during the COVID-19 pandemic, a unique period in human history.
The results reported in this study are highly connected to this unique context. 
For example, the speed of these two communities' explosion at the beginning is unusual and 
may not apply to other communities~\cite{nbcreddit}.
Although the generalization of insights from this work to 
highly-related communities in other contexts requires further investigation, we believe that it is valuable to bridge the literature on highly-related communities and crisis informatics in this unique context.

\para{Implications and future directions.}
Our observation that users in \rchinaflu and \rcoronavirus were similar at the early stage but diverge later on provides implications for community organizers. 
This suggests that when highly-related communities emerge, 
they may follow different directions and each develops its own identity.
In our example, even though \rchinaflu and \rcoronavirus are both for general discussions about COVID-19,
they attract people with different interests, and the separation grows over time.
Understanding the mechanism behind this phenomenon could be potentially useful for 
designing online spaces.

Our work also demonstrates that online communities do not only exist in the virtual world. 
User activity in online communities can be heavily embedded in the offline context.
COVID-related subreddits provide a unique opportunity for understanding the connections 
between the online and offline worlds in a crisis.
These communities only exist as a result of the COVID-19 pandemic. 
Many activity changes identified in their users are highly correlated
with how the epidemic unfolds in the real world.
Such observations emphasize the necessity of connecting online and offline data resources to 
explain online communities' dynamics and their relationships with on-going offline events.
An exciting future direction is to further understand the changes in online activities 
in the context of fine-grained offline events.

\section{Acknowledgement}
\label{sec:ack}

\thanks{We thank anonymous reviewers for their insightful comments, 
Daniel Strain from the CU Boulder Strategic Relations and Communications for his coverage and discussion of this work,
Scott Fredrick Holman from the CU Boulder Writing Center for his feedback 
and support during the writing process,
Jason Baumgartner for sharing the dataset that enabled this research.
This work is supported in part by the US National Science
Foundation (NSF) through grant IIS-1910225.}

\small
\bibliography{bibliography}

\end{document}